\documentclass[prb, twocolumn,tbtags,superscriptaddress,floatfix]{revtex4}
\usepackage{amsmath}
\usepackage{graphicx}
\usepackage{amssymb}
\usepackage{color,soul}
\usepackage[T1]{fontenc}
\usepackage{ae,aecompl}
\usepackage{natbib}

\begin{document}

\title{Low-frequency spectroscopy for quantum multi-level systems }
\date{\today }
\author{S.~N.~Shevchenko}
\affiliation{B. Verkin Institute for Low Temperature Physics and Engineering, Kharkov
61103, Ukraine}
\affiliation{V.~N. Karazin Kharkov National University, Kharkov 61022, Ukraine}
\affiliation{Theoretical Quantum Physics Laboratory, RIKEN Cluster for Pioneering
Research, Wako-shi, Saitama 351-0198, Japan}
\author{A. I. Ryzhov}
\affiliation{B. Verkin Institute for Low Temperature Physics and Engineering, Kharkov
61103, Ukraine}
\affiliation{V.~N. Karazin Kharkov National University, Kharkov 61022, Ukraine}
\author{Franco Nori}
\affiliation{Theoretical Quantum Physics Laboratory, RIKEN Cluster for Pioneering
Research, Wako-shi, Saitama 351-0198, Japan}
\affiliation{Physics Department, University of Michigan, Ann Arbor, MI 48109-1040, USA}

\begin{abstract}
A periodically driven quantum system with avoided level crossing experiences
both non-adiabatic transitions and wave-function phase changes. These result
in coherent interference fringes in the system's occupation probabilities.
For qubits, with repelling energy levels, such interference, named after
Landau-Zener-St\"{u}ckelberg-Majorana, displays arc-shaped resonance lines.
In the case of a multi-level system with an avoided level crossing of the
two lower levels, we demonstrate that the shape of the resonances can change
from convex arcs to concave heart-shaped and harp-shaped resonance lines.
Indeed, the whole energy spectrum determines the shape of such resonance
fringes and this also provides insight on the slow-frequency system
spectroscopy. As a particular example, we consider this for valley-orbit
silicon quantum dots, which are important for the emerging field of
valleytronics.
\end{abstract}

\maketitle

\section{Introduction}

Quantum systems can be reliably prepared, controlled, and probed. The
\textquotedblleft simplest nonsimple quantum problem\textquotedblright\ \cite%
{Berry95} is arguably a driven two-level system (a qubit), which can be used
for quantum sensing \cite{Degen17} and quantum information \cite{Buluta11}.
Due to the interplay of the non-adiabatic transitions between the energy
levels and the accumulation of the wave-function phase changes, the
interference fringes provide a convenient and powerful tool for controlling
and probing both the quantum system and its environment. This technique,
known as Landau-Zener-St\"{u}ckleberg-Majorana (LZSM) interferometry \cite%
{Shevchenko10}, is ubiquitously applied to\textit{\ two-level} quantum
systems. (For several experimental realizations in both superconducting and
semiconducting systems, see, e.g., Refs.~[%
\onlinecite{Oliver05, Sillanpaa06,
Wilson07, Izmalkov08, Sun09, Stehlik12, Gonzalez-Zalba16}].) However, a
generalization of this approach to\textit{\ multi-level} systems remains a
mostly open and topical subject, to which we devote the present work.

In order for LZSM physics to be directly relevant, a multi-level system has
to have a reasonable quasicrossing of the lower energy levels, also known as
avoided level crossing. Usually, multi-level systems have either all levels
coupled or all well separated. The former case contains transitions between
all energy levels and is known as amplitude spectroscopy \cite{Berns08,
Satanin12}. In the latter case, with a significant energy-level separation,
a slow drive would not produce non-adiabatic transitions due to negligibly
small tunneling probabilities, described by the Landau-Zener (LZ) formula.
The cure to this could be to \textquotedblleft dress\textquotedblright\ the
system with another, resonant, signal. Then, these conveniently prepared
dressed levels could be slowly driven and probed by means of LZSM physics.
This approach was demonstrated for superconducting qubits.\cite{Sun11,
Gong16} One message we would like to convey here is that a multi-level
system should be \textit{doubly driven}: by a \textit{resonant dressing}
signal and a \textit{slow} driving one. In different contexts, doubly-driven
quantum systems were studied in Refs.~[%
\onlinecite{Greenberg07, Greenberg08, Mefed99, Tuorila10, Silveri13,
Saiko14, Neilinger16}], while other examples of driven multi-level systems,
where LZSM physics is relevant, are Refs.~[%
\onlinecite{Chen11, Stehlik12, Kenmoe13, Ashhab16, Stehlik16, Sinitsyn17, Chatterjee18, Bogan18, Koski18, Gramajo18,
Parafilo18}].

So, our aim here is to consider how a multi-level system can be reduced to a
two-level one, being well separated from the upper ones but bearing
information about them. Here, instead of considering a general case, we
would rather focus on an example \cite{Berry95}: silicon double quantum dots
(DQDs) exploiting both orbital and valley degrees of freedom, which make
them multi-level systems \cite{Yang13, Burkard16, Zhao18, Mi18}. Such
systems present a unique opportunity of using the valley degree of freedom,
which is studied in the emerging field of valleytronics.\cite{Rozhkov17}

The rest of the paper is organized as follows. We will start in Sec.~II from
a four-state Hamiltonian for a silicon orbital-valley DQD, Ref.~[%
\onlinecite{Burkard16}]. (Another example of a four-state system is a device
with two coupled qubits, studied in Appendix~A.) We will discuss how to
prepare the DQD states for low-frequency LZSM\ spectroscopy by dressing them
with a resonant signal, with details presented in Appendix~B. The dressing
allows to reduce the four-level system to a two-level one. Then, in Sec.~III
we adopt the formulas from Ref.~[\onlinecite{Shevchenko10}] for this case.
In Sec.~IV we discuss the interference fringes obtained. We will also
analyze the shape of the resonant lines. For a generic dressed four-level
system, these are expected to be harp-shaped, which is demonstrated here for
the parameters used in the experiments in Ref.~[\onlinecite{Mi18}]. A
particular case, with a symmetric Hamiltonian, is analyzed in Appendix C. We
conclude with a discussion that these studies allow the means for
low-frequency spectroscopy for multi-level quantum systems.

\section{Bare and dressed energy levels}

Let us consider the four-state Hamiltonian for a silicon orbital-valley DQD~[%
\onlinecite{Burkard16}]:

\begin{equation}
H(t)=\!\left(
\begin{array}{cccc}
\frac{\epsilon (t)}{2}+E_{\mathrm{L}} & 0 & t_{\mathrm{d}} & t_{\mathrm{v}}
\\
0 & \frac{\epsilon (t)}{2} & -t_{\mathrm{v}} & t_{\mathrm{d}} \\
t_{\mathrm{d}} & -t_{\mathrm{v}} & -\frac{\epsilon (t)}{2}+E_{\mathrm{R}} & 0
\\
t_{\mathrm{v}} & t_{\mathrm{d}} & 0 & -\frac{\epsilon (t)}{2}%
\end{array}%
\right) =H_{0}+V_{\mathrm{d}}(t)  \label{Ham_QD}
\end{equation}%
with
\begin{equation}
V_{\mathrm{d}}(t)=\frac{1}{2}A_{\mathrm{d}}\sin \omega _{\mathrm{d}}t\sigma
_{z}^{(1)},
\end{equation}%
where $\sigma _{k}^{(1)}=\sigma _{k}\otimes \sigma _{0}$, and the $\sigma
_{k}$'s stand for the Pauli matrices. The $E_{\mathrm{L,R}}$ are the
left/right dot valley splittings, $t_{\mathrm{d}}$ and $t_{\mathrm{v}}$\ are
the inter-dot and inter-valley tunnel couplings, respectively. The energy
bias is chosen as%
\begin{equation}
\epsilon (t)=\varepsilon _{0}+A\sin \omega t+A_{\mathrm{d}}\sin \omega _{%
\mathrm{d}}t\equiv \varepsilon +A_{\mathrm{d}}\sin \omega _{\mathrm{d}}t%
\text{, \ \ \ }\omega \ll \omega _{\mathrm{d}},
\end{equation}%
which contains both the resonant dressing drive with frequency $\omega _{%
\mathrm{d}}$ and the slow spectroscopy drive $\varepsilon =\varepsilon
_{0}+A\sin \omega t$ with frequency $\omega \ll \omega _{\mathrm{d}}$. Our
approach consists of two steps. In the first step (\textquotedblleft
dressing\textquotedblright ), we will ignore the slow signal and consider $%
\varepsilon $ to be a time-independent value. We will demonstrate how to
reduce this system to a two-level one. (For other similar cases, when a
multi-level structure is reduced to a two-level system see Refs.~[%
\onlinecite{Qi17, Pietikainen17b}].) After incorporating this fast drive as
the dressing, we will then add\ the slow time dependence, contained in the
variable $\varepsilon $.

\begin{figure}[t]
\includegraphics[width=8.5 cm]{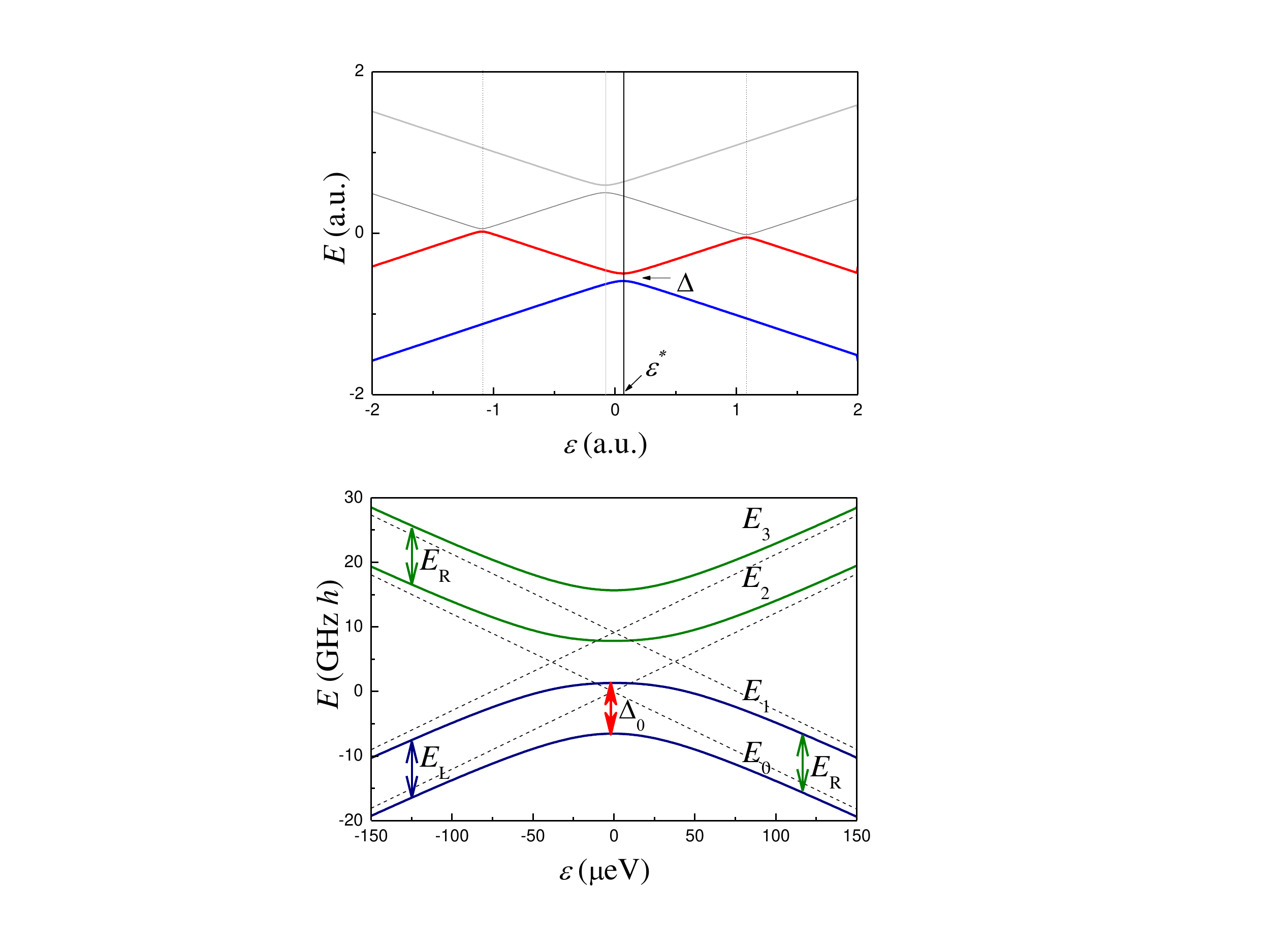}
\caption{\textbf{A four-level system: Energy levels of an orbital-valley DQD.%
} The four energy levels in the absence of tunneling, $t_{\mathrm{d}}=t_{%
\mathrm{v}}=0$, shown by the dashed lines, experience four crossings. In
general case, shown by the solid lines, the degeneracy is lifted, and the
four energy levels $E_{i}$ are plotted for the parameters described in the
text.}
\label{Fig1}
\end{figure}

Consider first the energy levels of our four-level system. These are the
eigenstates of the Hamiltonian $H_{0}$. In the absence of tunneling, $t_{%
\mathrm{d}}=t_{\mathrm{v}}=0$, these are given by the diagonal matrix
elements in Eq.~(\ref{Ham_QD}). These are the four straight intersecting
lines in Fig.~\ref{Fig1}. Non-zero tunneling lifts the degeneracies. For
calculations in this work we choose the parameters for a silicon
orbital-valley DQD from Ref.~[\onlinecite{Mi18}]: $E_{\mathrm{L}}=37.5~\mu
\mathrm{eV}$, $E_{\mathrm{R}}=38.3~\mu \mathrm{eV}$, $t_{\mathrm{d}%
}=25.4~\mu \mathrm{eV}$,\ $t_{\mathrm{v}}=11.8~\mu \mathrm{eV}$. (Another
possible realization of a four-level structure, describing a two-qubit
system, is given in Appendix A.) The spectrum with these parameters is shown
in Fig.~\ref{Fig1}. The chosen parameters, which enter the Hamiltonian~(\ref%
{Ham_QD}), result in the minimal energy difference $\Delta
_{0}=(E_{1}-E_{0})_{\min }=7.845~$GHz$\cdot h$, and this takes place at very
small offset, $\varepsilon =\varepsilon ^{\ast }=-4\cdot 10^{-3}~$GHz.
(Since we use both energy and frequency units, we note, for convenience,
that $1~\mu \mathrm{eV}=0.2418~$GHz$\cdot h$.) Such large splitting $\Delta
_{0}$ does not allow low-frequency spectroscopy because, according to the
adiabatic theorem and the LZ formula, there would be no excitation for
low-frequency driving. So, we will first \textquotedblleft
dress\textquotedblright\ the \textquotedblleft bare\textquotedblright\
spectrum in Fig.~\ref{Fig1}.

Accordingly, consider now the resonant driving with $\epsilon
(t)=\varepsilon +A_{\mathrm{d}}\sin \omega _{\mathrm{d}}t$ and $\hbar \omega
_{\mathrm{d}}\sim \Delta _{0}$. The detailed procedure is described in
Appendix B. This results in the shift of the energy levels and the
separation of the lower two levels from the upper ones. These become $%
\widetilde{E}_{0,1}=E_{0,1}\pm \hbar \omega _{\mathrm{d}}/2$. What matters
for the low-frequency evolution then is the distance between these
meaningful energy levels,
\begin{equation}
\Delta \widetilde{E}=\Delta E-\hbar \omega _{\mathrm{d}},
\end{equation}%
where $\Delta \widetilde{E}=\widetilde{E}_{1}-\widetilde{E}_{0}$ and $\Delta
E=E_{1}-E_{0}$. Thus, we have mapped a multi-level system into a \textit{%
two-level dressed} one.

\begin{widetext}

\begin{figure}[t]
\includegraphics[width=8.5 cm]{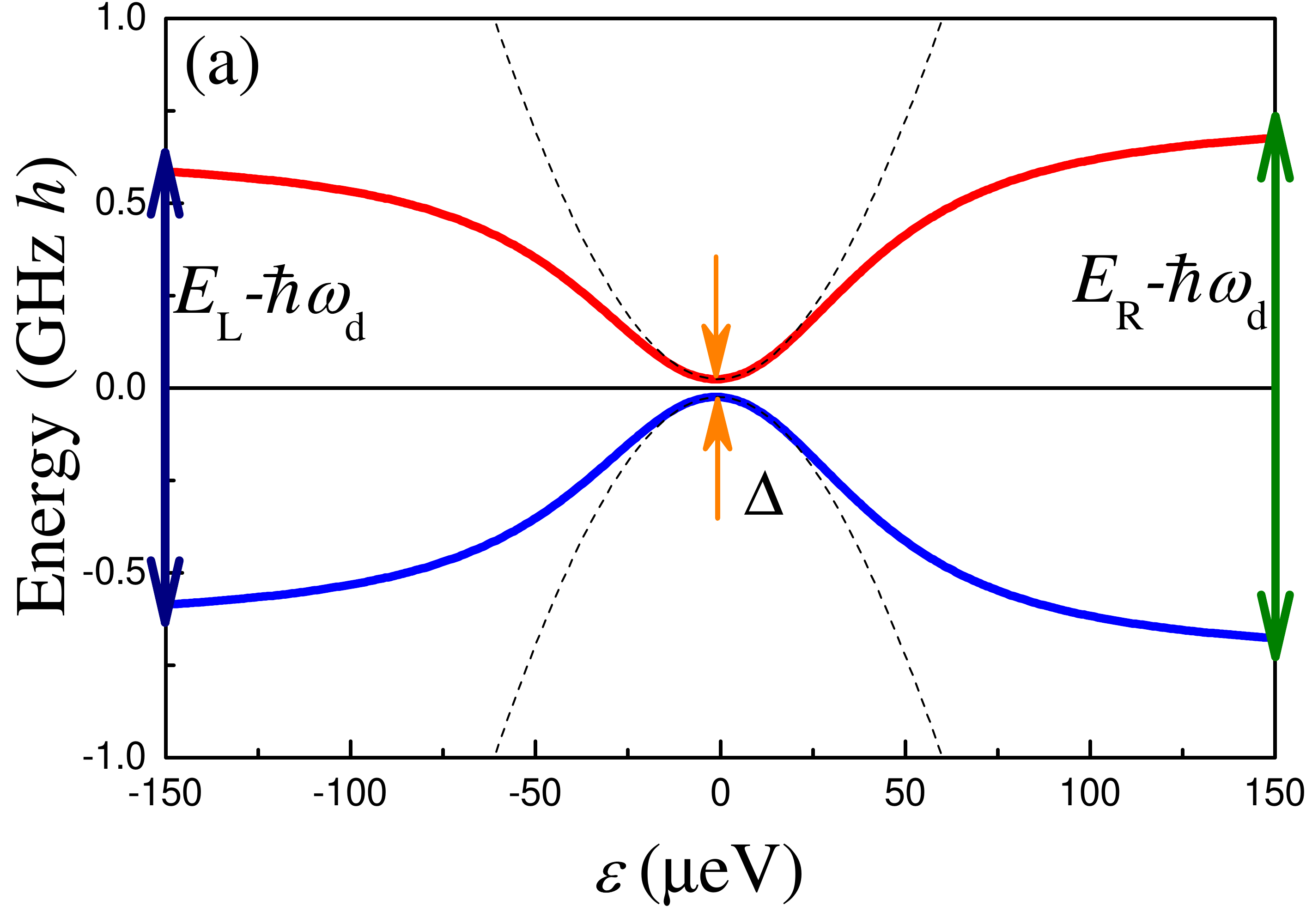}
\includegraphics[width=8.5 cm]{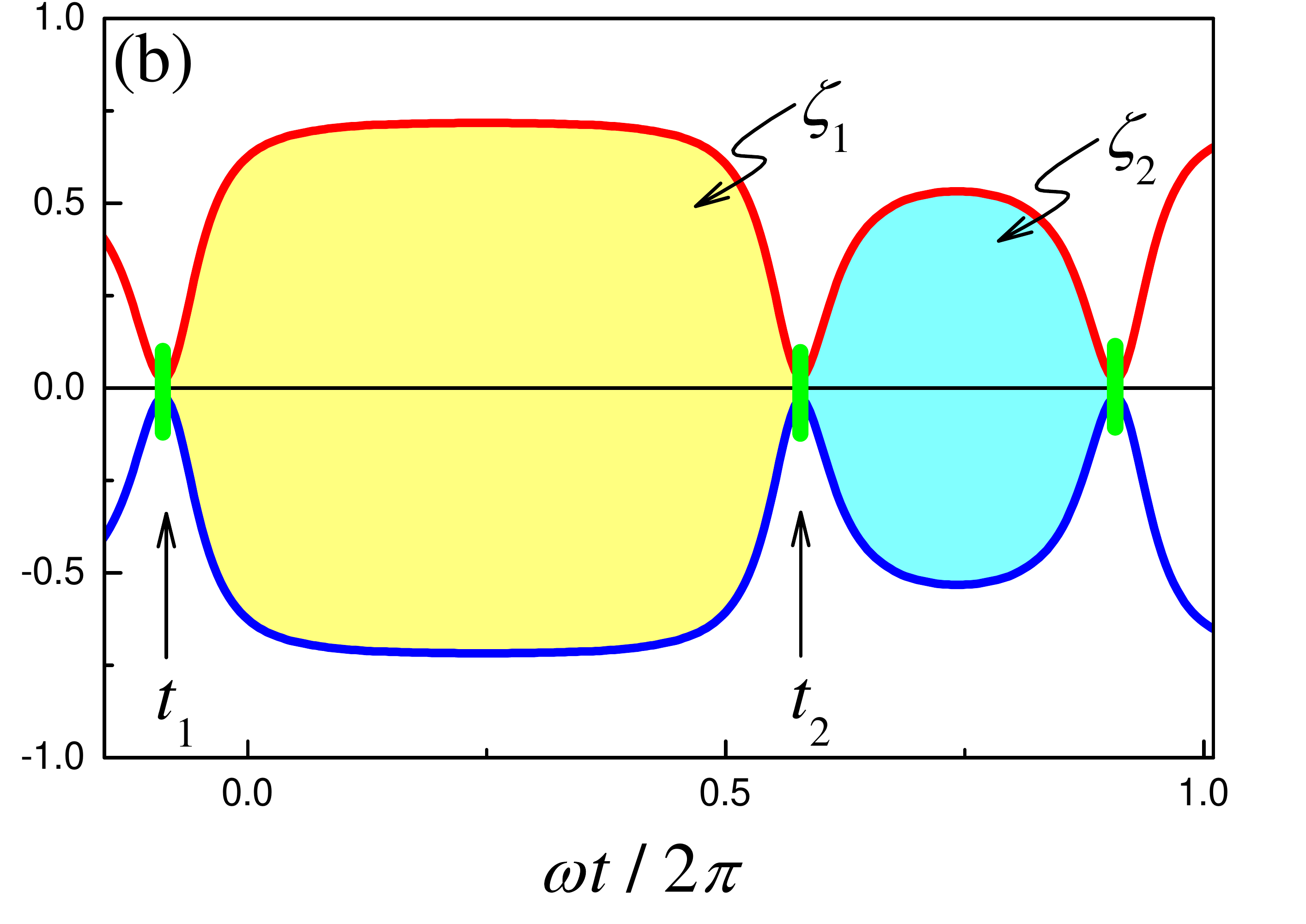}
\caption{\textbf{Dressed energy levels.} In (a) the two lowest dressed
states are shown by plotting $\pm \Delta \widetilde{E}/2$ versus $\protect%
\varepsilon $, where $\Delta \widetilde{E}=\Delta E-\hbar \protect\omega _{%
\mathrm{d}}$ and $\Delta E=E_{1}-E_{0}$. The distance between the dressed
energy levels shows the avoided-level crossing $\Delta $ at around $%
\protect\varepsilon =0$ and tends to $E_{\mathrm{L,R}}-\hbar \protect\omega %
_{\mathrm{d}}$ to the left and right, respectively. The dashed parabolas
correspond to the qubit-like spectrum at small $\protect\varepsilon $. In (b)
the same dressed energy levels are shown versus time for $\protect%
\varepsilon =\protect\varepsilon _{0}+A\sin \protect\omega t$, with $\protect%
\varepsilon _{0}=50~\protect\mu \mathrm{eV}$ and $A=100~\protect\mu \mathrm{%
eV}$. Several avoided-level crossings take place at $t=t_{1}$, $t_{2}$, and $%
t_{2}+2\protect\pi /\protect\omega $. The probabilistic LZ transitions are
shown in these points by thick vertical green dashes. Between these
avoided-level crossings, the wave-function phases $\protect\zeta _{1,2}$ are
accumulated. These phases, equal to the areas under the energy-level curves,
result in observable interference fringes.}
\label{Fig2}

\end{figure}

\end{widetext}

To better compare with qubits, it is instructive to plot the equivalent
(mirror-reflected) energy levels, $\pm \Delta \widetilde{E}/2$, with the
same distance $\Delta \widetilde{E}$, instead of $\widetilde{E}_{0,1}$. The
driving frequency should be taken close to $\Delta _{0}$, and then with $%
\omega _{\mathrm{d}}/2\pi =7.796~$GHz of [\onlinecite{Mi18}], we have the
dressed avoided level distance $\Delta =\Delta _{0}-\hbar \omega _{\mathrm{d}%
}=0.049~$GHz$\cdot h$. The dressed energy levels, featuring this avoided
level crossing, are shown in Fig.~\ref{Fig2}(a) as a function of the energy
bias $\varepsilon $.

Close to the avoided level crossing, we can expand $\left( \Delta E\right)
^{2}$ in series in $\varepsilon $ and obtain $\Delta \widetilde{E}=\sqrt{%
\Delta _{0}^{2}+0.16\varepsilon ^{2}}-\hbar \omega _{\mathrm{d}}$. The
respective curves are shown by the dashed lines in Fig.~\ref{Fig2}(a). This
formula is useful for the description of the dynamics with $\varepsilon <E_{%
\mathrm{L,R}}$.

Hereafter, the slow signal, driving the qubit, will be taken with $\omega
\sim \Delta $, so that to have a non-trivial LZ probability, $P_{\mathrm{LZ}%
}\sim 1$. Then $\varepsilon =\varepsilon _{0}+A\sin \omega t$ describes the
low-frequency parametric time dependence of the energy levels. Imagine that
we start at $\varepsilon =\varepsilon _{0}$ with, say, $\varepsilon
_{0}=50~\mu \mathrm{eV}$ in Fig.~\ref{Fig2}(a). Then the dynamics
corresponds to first increasing the bias $\varepsilon $ up to $\varepsilon
=\varepsilon _{0}+A$, and then decreasing it to $\varepsilon =\varepsilon
_{0}-A$. Respectively, the energy levels $\pm \frac{1}{2}\Delta \widetilde{E}%
(\varepsilon )$ will change, as shown in Fig.~\ref{Fig2}(b). Each time the
system passes through $\varepsilon =0$ in Fig.~\ref{Fig2}(a), we have the
avoided level crossing in Fig.~\ref{Fig2}(b). Such dynamics is described by
the so-called \textit{adiabatic-impulse} model, as detailed in Refs.~[%
\onlinecite{Ashhab07, Shevchenko10}] and references therein. This model
combines both intuitive clarity and quantitative accuracy. So, we devote the
next section to this.

\section{LZSM for a multi-level system}

We now would like to calculate the occupation probabilities for the
two-level system with the energy levels $\pm \frac{1}{2}\Delta \widetilde{E}%
(\varepsilon )$, shown in Fig.~\ref{Fig2}. The adiabatic-impulse model
considers the dynamics to be adiabatic, when far from the avoided level
crossings, with non-adiabatic transitions at the points of minimal
energy-level distance. The former stages are described by the accumulation
of the wave-function phases, while the latter are characterized by the LZ
transition formula. With this we can generalize the formulas for the
slow-passage case from Refs.~[\onlinecite{Shevchenko10, Shevchenko12}],
giving the upper-level time-averaged occupation probability%
\begin{equation}
P_{+}=\frac{P_{\mathrm{LZ}}(1+\cos \zeta _{+}\cos \zeta _{-})}{\sin
^{2}\zeta _{+}+2P_{\mathrm{LZ}}(1+\cos \zeta _{+}\cos \zeta _{-})},
\label{Pp_with_offset}
\end{equation}%
where
\begin{eqnarray}
\zeta _{+} &=&\zeta _{1}+\zeta _{2}+\varphi ,\text{ }\zeta _{-}=\zeta
_{1}-\zeta _{2},\text{ } \\
\zeta _{1} &=&\frac{1}{2\hbar }\int\limits_{t_{1}}^{t_{2}}\Delta \widetilde{E%
}(t)dt,\text{ }\zeta _{2}=\frac{1}{2\hbar }\int_{t_{2}}^{t_{1}+2\pi /\omega
}\Delta \widetilde{E}(t)\;dt,  \notag \\
\Delta \widetilde{E} &=&E_{1}-E_{0}-\hbar \omega _{\mathrm{d}},\text{ }%
\varepsilon =\varepsilon _{0}+A\sin \omega t,  \notag \\
\omega t_{1} &=&\text{asin}\left( -\frac{\varepsilon _{0}}{A}\right) ,\text{
}\omega t_{2}=\pi -\omega t_{1},  \notag \\
\varphi &=&-\frac{\pi }{2}+2\delta (\ln \delta -1)+2\arg \Gamma (1-i\delta ),
\notag \\
\delta &=&\frac{\Delta ^{2}}{4v},\text{ }v=A\hbar \omega \sqrt{1-\left(
\frac{\varepsilon _{0}}{A}\right) ^{2}}.  \notag
\end{eqnarray}%
And the probability of the non-adiabatic transition to the upper adiabatic
level during the avoided level passage is given by the Landau-Zener formula $%
P_{\mathrm{LZ}}=\exp \left( -2\pi \delta \right) $. Here $\Gamma $ denotes
the gamma function. Note that for sufficiently small frequency ($\delta \gg
1 $) one could assume $\varphi \approx -\pi $, though in the equation above
we keep the complete form of the phase, for the sake of generality.

\begin{figure}[t]
\includegraphics[width=8.5 cm]{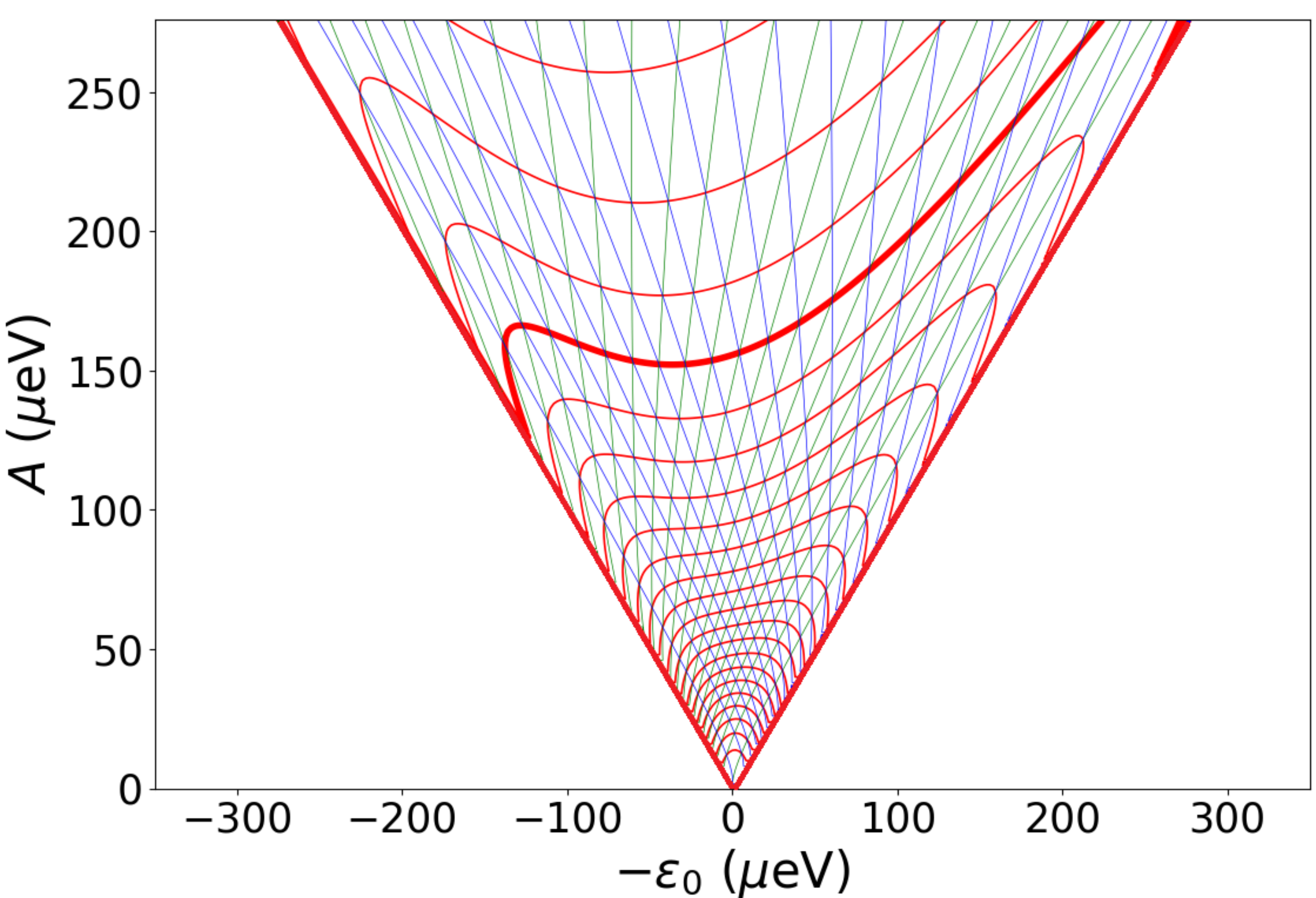}
\caption{\textbf{Harp-shaped resonance fringes and nodes.} The position of
the resonances is shown in red for different values of $k$, Eq.~(\protect\ref%
{resonances}). These curves are equidistant arcs for small $k^{\prime }$s
and harp-shaped lines with increasing distance for higher $k^{\prime }$s.
When both phases $\protect\zeta _{1,2}/\protect\pi $ equal to half-integer
numbers [Eq.~(\protect\ref{nodes})], the resonances are suppressed, and
their positions are given by the intersection of the green and blue lines. }
\label{Fig3}
\end{figure}

Formula~(\ref{Pp_with_offset}) defines the lines (arcs), along which the
resonances are situated:%
\begin{equation}
\zeta _{+}=k\pi .  \label{resonances}
\end{equation}%
Under this condition, the upper-level occupation probability becomes the
highest possible, $P_{+}=1/2$. The width of the resonance lines is defined
by the numerator in Eq.~(\ref{Pp_with_offset}), which tends to zero when%
\begin{equation}
\zeta _{1}=\frac{\pi }{2}+l\pi \text{ \ and \ \ }\zeta _{2}=\frac{\pi }{2}%
+m\pi ,  \label{nodes}
\end{equation}%
where $l$ and $m$ are integers. Note that these intersect at $\zeta
_{1}+\zeta _{2}=(l+m+1)\pi \equiv k\pi $, which means that the nodes are
situated on the resonance lines, defined by Eq.~(\ref{resonances}). These
are plotted in Fig.~\ref{Fig3} for $\omega /2\pi =50$ MHz. The resonance
line with $k=20$ is shown bolder in Fig.~\ref{Fig3} to show that these are
the harp-shaped resonance lines with convex shapes. Such harp-shaped
resonances were reported recently in Ref.~[\onlinecite{Mi18}].

\section{Discussion: relevance for low-frequency spectroscopy}

The positions of the resonances in Fig.~\ref{Fig3} bear information about
the initial four-state Hamiltonian. Thus, these observations could be used
for defining the system parameters, which effectively correspond to the
spectroscopy of a multi-level system. Let us now summarize several
distinctive features, which could be useful for this type of spectroscopy.

\begin{figure}[t]
\includegraphics[width=8.5 cm]{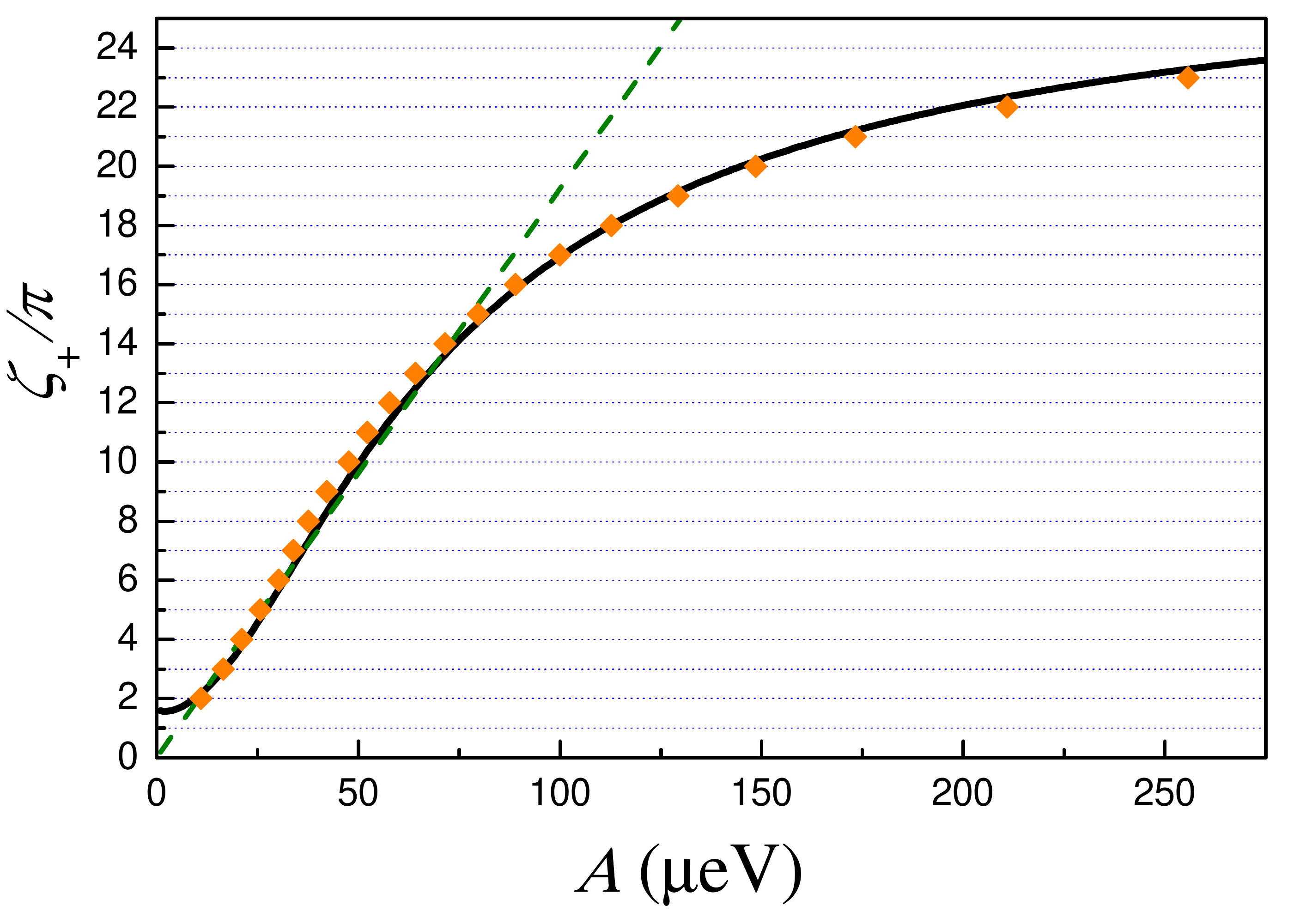}
\caption{\textbf{Power dependence of the resonances at zero offset, }$%
\protect\varepsilon _{0}=0$\textbf{. }The solid black curve is given by $%
\protect\zeta _{+}$. The position of the resonances is defined by the
equation $\protect\zeta _{+}=k\protect\pi $ and this corresponds to integer
parts of $\protect\zeta _{+}/\protect\pi $, which are marked as horizontal
grid lines. At low bias, i.e. for $k\leq 15$, these can be fit by the
constant-slope green dashed line. This means that the resonance arcs are
approximately equidistant. At higher driving power $A$, the inter-resonance
distance monotonically increases. The orange diamonds correspond to the
resonances from the experimental data in Fig.~2(a) of Ref.~[\onlinecite{Mi18}%
].}
\label{Fig4}
\end{figure}

\begin{itemize}
\item The resonances are limited by the inclined lines, in the region $%
A>\left\vert \varepsilon _{0}-\varepsilon ^{\ast }\right\vert $. This is
because otherwise the avoided level crossing is not reached and there is no
transition from the ground state to the excited one. The inclination of
these lines could be useful for power calibration.

\item For small driving amplitudes, $A<E_{\mathrm{L,R}}$, we have a
qubit-like spectrum, and accordingly, the arcs are equidistant and symmetric.

\item With increasing the driving amplitude, starting at $A\sim E_{\mathrm{%
L,R}}$, the resonances become asymmetric.

\item With further increasing the driving amplitude, the shape of the
resonance lines changes from convex to concave, producing harp-shaped
curves. This is because the energy-level distance changes from increasing to
becoming constant, see Fig.~\ref{Fig2}. In the symmetric case, with $E_{%
\mathrm{L}}=E_{\mathrm{R}}$, the curves are symmetric, and this case is
analyzed in Appendix~C.

\item At large driving power, the resonance lines are increasingly
separated. This can be conveniently studied along the line $\varepsilon
_{0}=0$ in Fig.~\ref{Fig3}. This is done in Fig.~\ref{Fig4}. There, one can
see the equidistant resonance position at smaller driving power $A$, as
described by the inclined dashed line, and the increasing inter-resonance
separation at larger $A$.
\end{itemize}

Our calculations are related to the experimental parameters of Ref.~[%
\onlinecite{Mi18}]. In particular, note the good agreement shown in Fig.~\ref%
{Fig4}. Moreover, our general approach can be applied to \textit{any} other
\textit{multi-level} system. Our general formulation allows the flexible
application to other systems, easy numerical calculations, as well as
analytical analysis in various limiting cases. These are not possible by a
direct numerical solution, without our more-analytical approach.

\section{Conclusion}

We have demonstrated how a multi-level system could be reduced to a
two-level one, by applying a resonant dressing signal. The obtained
two-level system is remarkably distinct from a qubit because at larger bias
the energy levels become equally separated, and not repelling. This
distinction results in that the resonance fringes follow harp-shaped lines.
Since the dressed two levels bear information about the initial multi-state
system, the unusual and versatile properties of such interferometric
features could be adopted for multi-level systems' spectroscopy.

\begin{acknowledgments}
We thank M.~F.~Gonzalez-Zalba and K. Ono for useful and stimulating
discussions and J.~R.~Petta for sharing with us the experimental results of
Ref.~[\onlinecite{Mi18}]\ prior to publication. F.N. is supported in part by
the MURI Center for Dynamic Magneto-Optics via the Air Force Office of
Scientific Research (AFOSR) (FA9550-14-1-0040), Army Research Office (ARO)
(Grant No. W911NF-18-1-0358), Asian Office of Aerospace Research and
Development (AOARD) (Grant No. FA2386-18-1-4045), Japan Science and
Technology Agency (JST) (Q-LEAP program, ImPACT program and CREST Grant No.
JPMJCR1676), Japan Society for the Promotion of Science (JSPS) (JSPS-RFBR
Grant No. 17-52-50023, and JSPS-FWO Grant No. VS.059.18N), RIKEN-AIST
Challenge Research Fund, and the John Templeton Foundation.
\end{acknowledgments}

\appendix

\section{A two-qubit four-level system}

\begin{figure}[t]
\includegraphics[width=8cm]{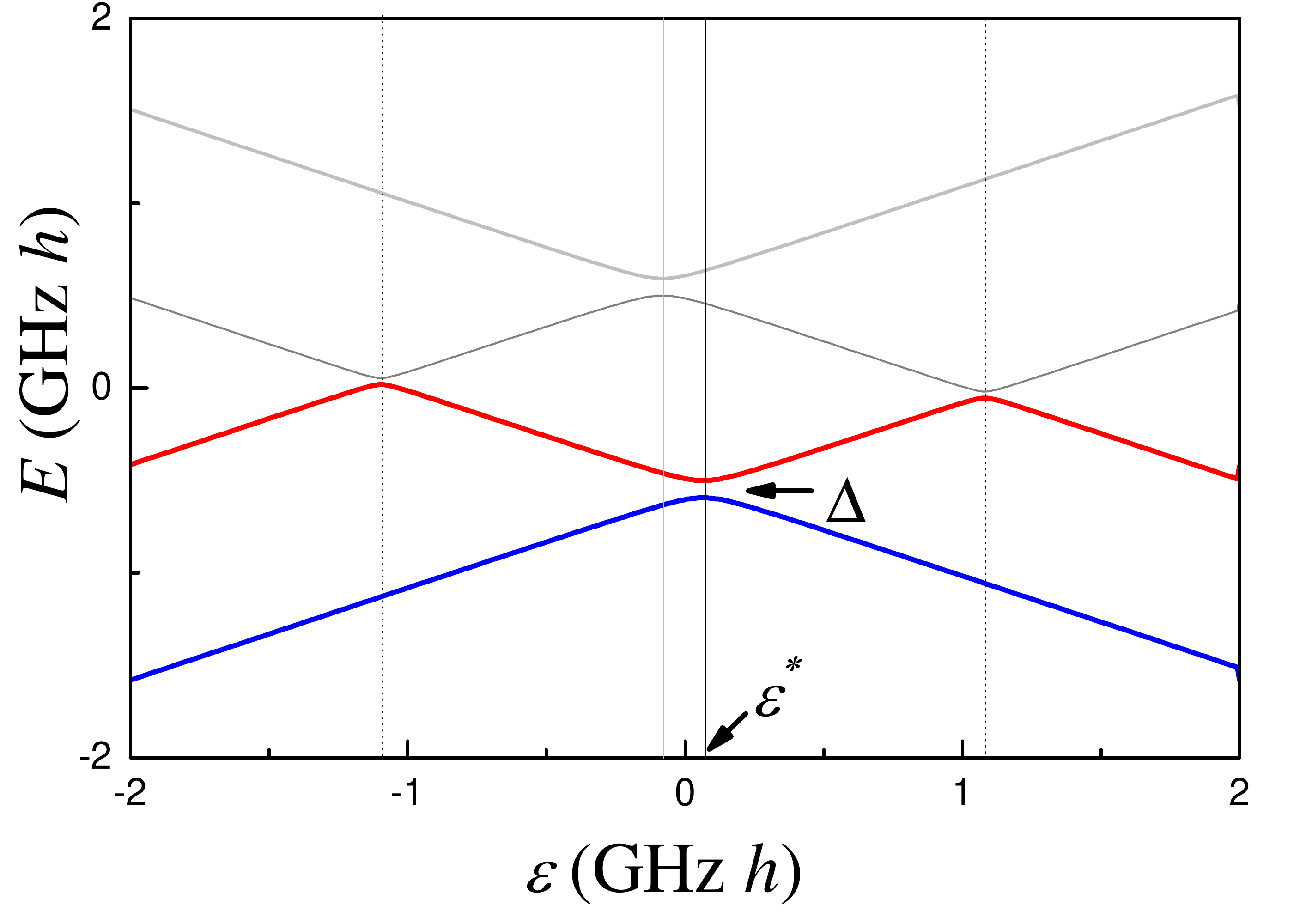}
\caption{\textbf{Energy levels of a two-qubit system.} These are shown as an
example of the situation where the lower two energy levels have a small
avoided level crossing $\Delta $ and the distance between the energy levels
increases at around this point and tends to constants at high values of the
bias $\protect\varepsilon $.}
\label{FigA1}
\end{figure}

While multi-level quantum systems could be found in different contexts, we
would like to present one additional example: a system of two coupled
qubits. Let us now consider the Hamiltonian \cite{Denisenko10, Temchenko11,
Gramajo17}

\begin{gather}
H=-\frac{1}{2}\sum_{i=1,2}\left( \Delta _{i}\sigma _{x}^{(i)}+\varepsilon
_{i}\sigma _{z}^{(i)}\right) +\frac{J}{2}\sigma _{z}^{(1)}\sigma _{z}^{(2)}=
\label{H0} \\
=-\frac{1}{2}\!\left(
\begin{array}{cccc}
\!\varepsilon _{1}\!+\!\varepsilon _{2}\!-\!J\! & \Delta _{2} & \Delta _{1}
& 0 \\
\Delta _{2} & \!\varepsilon _{1}\!-\!\varepsilon _{2}\!+\!J\! & 0 & \Delta
_{1} \\
\Delta _{1} & 0 & \!-\!\varepsilon _{1}\!+\!\varepsilon _{2}\!+\!J\! &
\Delta _{2} \\
0 & \Delta _{1} & \Delta _{2} & \!-\!\varepsilon _{1}\!-\!\varepsilon
_{2}\!-\!J\!%
\end{array}%
\right) ,  \notag
\end{gather}%
where $\sigma _{k}^{(1)}=\sigma _{k}\otimes \sigma _{0}$ and $\sigma
_{k}^{(2)}=\sigma _{0}\otimes \sigma _{k}$. Let us choose $\varepsilon _{2}$
to be a constant and $\varepsilon _{1}$ to have an alternating value: $%
\varepsilon _{2}=J$ (just for simplification) and
\begin{equation}
\varepsilon _{1}\equiv \varepsilon =\varepsilon _{0}+A\sin \omega t.
\end{equation}%
This would make the Hamiltonian somewhat resembling the one in Eq.~(\ref%
{Ham_QD}). Then the Hamiltonian becomes

\begin{equation}
H=-\frac{1}{2}\!\left(
\begin{array}{cccc}
\varepsilon & \Delta _{2} & \Delta _{1} & 0 \\
\Delta _{2} & \varepsilon & 0 & \Delta _{1} \\
\Delta _{1} & 0 & -\varepsilon +2J & \Delta _{2} \\
0 & \Delta _{1} & \Delta _{2} & -\varepsilon -2J%
\end{array}%
\right) =H_{0}+V_{\mathrm{d}}(t)  \label{H_2qbs}
\end{equation}%
with
\begin{equation}
V_{\mathrm{d}}(t)=-\frac{1}{2}A\sin \omega t\sigma _{z}^{(1)}.
\end{equation}%
In Fig.~\ref{FigA1} we choose: $\Delta _{\mathrm{1}}/h=0.1$ GHz, $\Delta _{%
\mathrm{2}}/h=1$ GHz, and $J/h=0.2$ GHz. Such parameters give the minimal
splitting $\Delta /h=73$ MHz and the shift $\varepsilon ^{\ast }/h=93$ MHz.
The lowest eigenvalues of $H_{0}$, denoted by $E_{0}$ and $E_{1}$, are shown
as the red and blue curves in Fig.~\ref{FigA1}.

Note that for a two-qubit four-level system, the energy levels are similar
to the ones presented in Fig.~\ref{Fig2}(a), in that they have small avoided
level crossing, an increasing energy-level distance for small bias, and a
constant distance for higher bias. An important distinction is that these
bare levels are \textit{not} separated from the upper ones. Transitions to
the upper states would produce additional interference fringes like in
Refs.~[\onlinecite{Berns08, Satanin12}].

\section{Dressing}

In this Appendix we consider how the resonantly driven four-state DQD can be
reduced to a dressed two-level system. We start from the time-dependent
Hamiltonian, Eq.~(\ref{Ham_QD}) with $\epsilon (t)=\varepsilon +A_{\mathrm{d}%
}\sin \omega _{\mathrm{d}}t$, with $\varepsilon $ assumed here being
time-independent, $H_{0}$ corresponds to $\epsilon (t)\rightarrow
\varepsilon $, and
\begin{equation}
V_{\mathrm{d}}(t)=\frac{1}{2}A_{\mathrm{d}}\cos \left( \omega _{\mathrm{d}%
}t\right) \sigma _{z}^{(1)}
\end{equation}%
with $\sigma _{z}^{(1)}=\sigma _{z}\otimes \sigma _{0}$. The stationary
Hamiltonian $H_{0}$ is diagonalized by the matrix $S$ (which can be found
numerically):
\begin{equation}
S^{\dag }H_{0}S=H_{0}^{\prime }=\mathrm{diag}(E_{0},E_{1},E_{2},E_{3}).
\end{equation}%
Then, the same procedure should be done with $V_{\mathrm{d}}(t)$; we denote
the matrix $V=S^{\dag }\sigma _{z}^{(1)}S$. And then, similarly to how this
is done for qubits, e.g.~in Ref.~[\onlinecite{Shevchenko14}], we make the
unitary transformation $U=\exp \left( i\omega _{\mathrm{d}}t\sigma
_{z}^{(2)}/2\right) $ and omit the fast-rotating terms, which means the
rotating-wave approximation. We obtain the Hamiltonian of the dressed DQD:%
\begin{eqnarray}
\widetilde{H} &=&\left(
\begin{array}{cccc}
E_{0}+\frac{\hbar \omega _{\mathrm{d}}}{2} &  &  & 0 \\
& E_{1}-\frac{\hbar \omega _{\mathrm{d}}}{2} &  &  \\
&  & E_{2}+\frac{\hbar \omega _{\mathrm{d}}}{2} &  \\
0 &  &  & E_{3}-\frac{\hbar \omega _{\mathrm{d}}}{2}%
\end{array}%
\right)  \notag \\
&&+\frac{A_{\mathrm{d}}}{4}\left(
\begin{array}{cccc}
0 & 0 & V_{02} & V_{03} \\
0 & 0 & V_{12} & V_{13} \\
V_{20} & V_{21} & 0 & 0 \\
V_{30} & V_{31} & 0 & 0%
\end{array}%
\right) .  \label{dressed_DQD}
\end{eqnarray}%
Here $V_{ij}$ are the elements of the matrix $V$.

\begin{figure}[t]
\includegraphics[width=8cm]{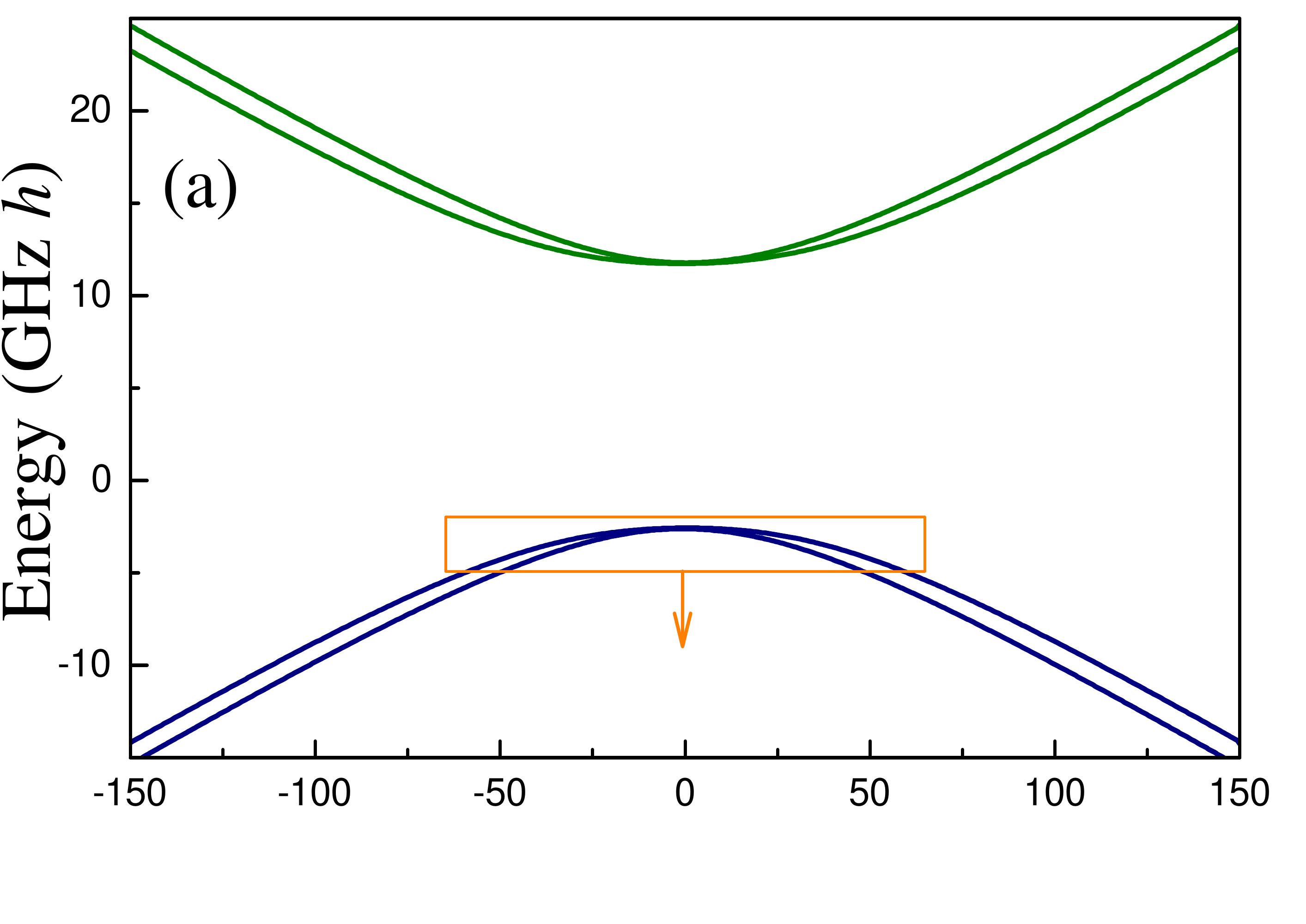} \includegraphics[width=8cm]{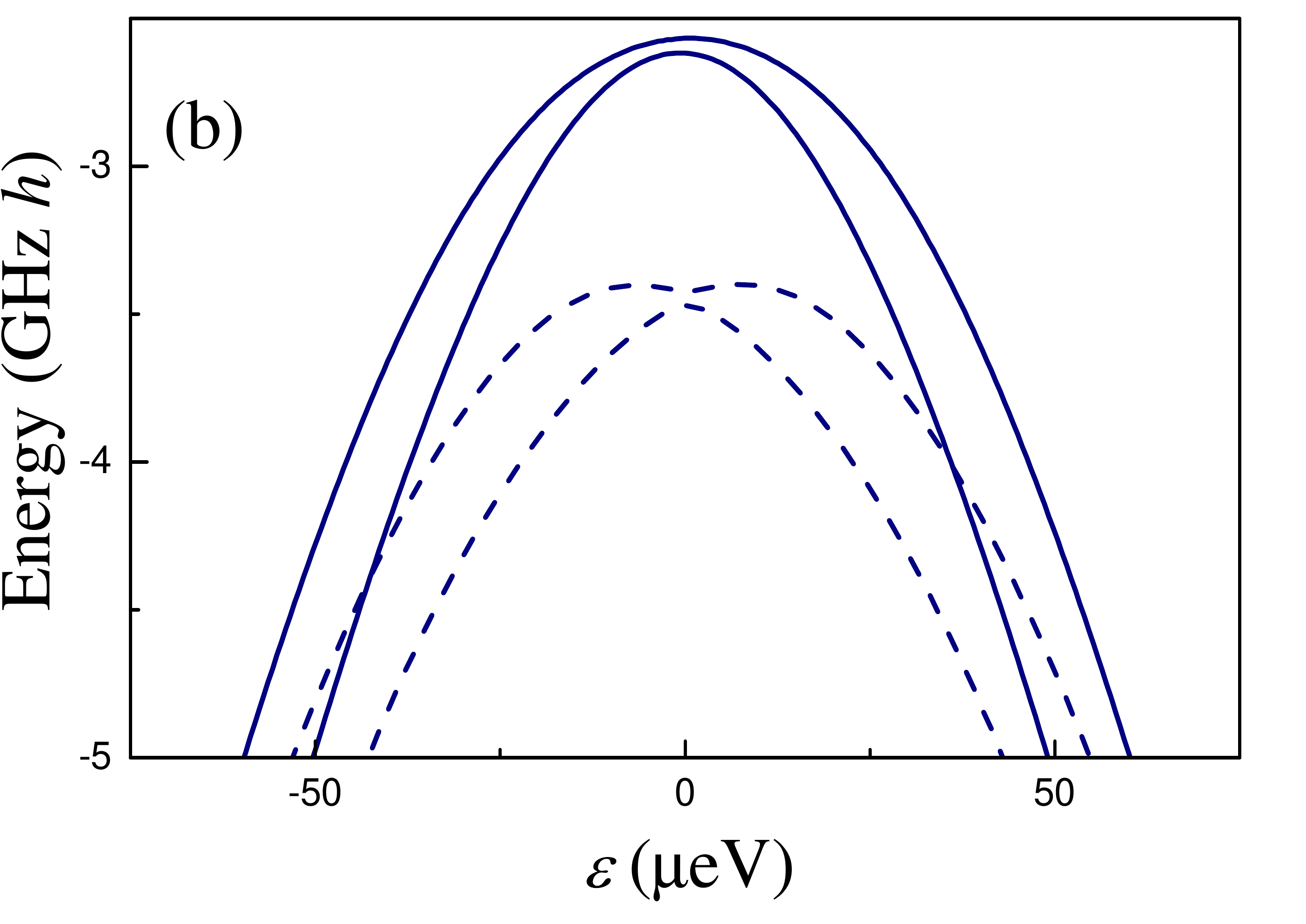}
\caption{\textbf{Energy levels of the dressed DQD.} (a)~At low driving
power, $A_{\mathrm{d}}\rightarrow 0$, these are defined by $E_{i}\pm \hbar
\protect\omega _{\mathrm{d}}/2$, see Eq.~(\protect\ref{dressed_DQD}).
(b)~Close-up of the lowest two dressed energy levels. The solid lines are
the same as in (a), for low driving power. This displays the minimal
energy-level distance at around $\protect\varepsilon _{0}=0$ given by $%
\Delta =(E_{1}-E_{0})_{\min }-\hbar \protect\omega _{\mathrm{d}}$. The
dashed lines show how the driving changes the dressed energy levels; these
are plotted for $A_{\mathrm{d}}/\Delta =2$. }
\label{FigB1}
\end{figure}

If we neglect the driving amplitude, $A_{\mathrm{d}}\rightarrow 0$, the
dressed energy levels are given by the shifted bare ones: $\widetilde{E}%
_{i}=E_{i}\pm \hbar \omega _{\mathrm{d}}$. These are plotted in Fig.~\ref%
{FigB1}(a). The effect of the driving, for non-zero $A_{\mathrm{d}}$, is
shown in Fig.~\ref{FigB1}(b) for the lowest two levels and $A_{\mathrm{d}%
}/\Delta =2$.

Figure~\ref{FigB1}(a) demonstrates that the lowest two dressed states are
well isolated from the upper ones. This allows to limit the description of
our system to two levels only,
\begin{equation}
\widetilde{E}_{0,1}=E_{0,1}\pm \frac{\hbar \omega _{\mathrm{d}}}{2}.
\end{equation}%
The distance between these levels is
\begin{equation}
\Delta \widetilde{E}=\Delta E-\hbar \omega _{\mathrm{d}},
\end{equation}%
where $\Delta \widetilde{E}=\widetilde{E}_{1}-\widetilde{E}_{0}$ and $\Delta
E=E_{1}-E_{0}$.

\section{Heart-shaped (concave) resonance fringes}

\begin{figure}[t]
\includegraphics[width=8cm]{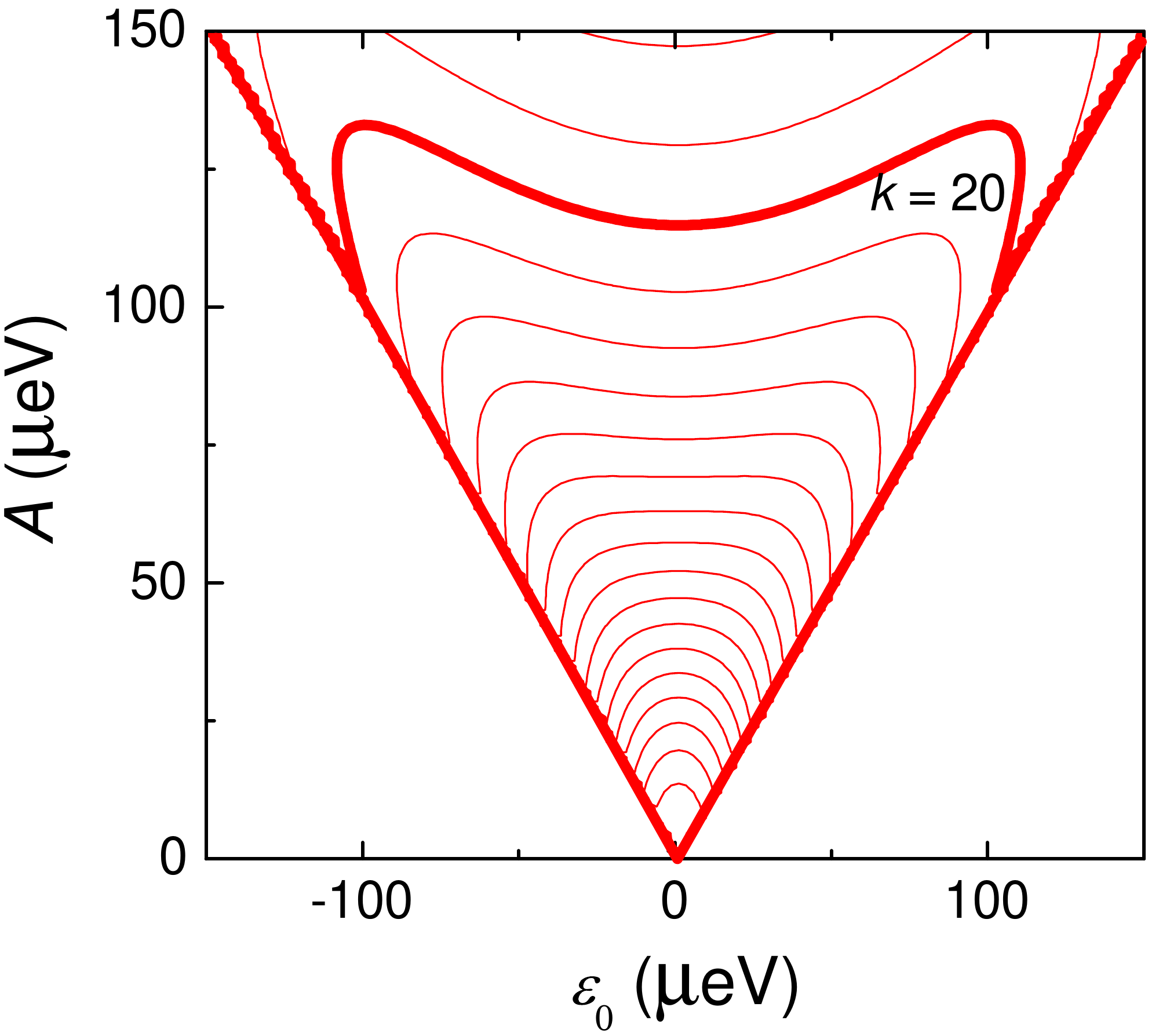}
\caption{\textbf{Resonances for a symmetric DQD.} These are calculated
analogously to the resonances shown in Fig.$~$\protect\ref{Fig3} with the
only difference that $E_{\mathrm{L}}=E_{\mathrm{R}}=38.3~\protect\mu $eV.
Resonances with $k$ from $4$ to $22$ are shown. We make bolder the resonance
line with $k=20$ to demonstrate that these are heart-shaped resonance lines
with the convex shape at $\protect\varepsilon _{0}=0$.}
\label{Fig:heart}
\end{figure}

Consider a symmetric DQD, being the same orbital-valley one described by
Eq.~(\ref{Ham_QD}), with only one distinction that now we assume%
\begin{equation}
E_{\mathrm{L}}=E_{\mathrm{R}}.
\end{equation}%
With this simplification, we can obtain expressions for the four energy
levels%
\begin{equation}
E_{0,1,2,3}=\frac{E_{\mathrm{R}}}{2}\pm \left\{ \left[ \frac{E_{\mathrm{R}}}{%
2}\pm \sqrt{\left( \frac{\varepsilon }{2}\right) ^{2}+t_{\mathrm{d}}^{2}}%
\right] ^{2}+t_{\mathrm{v}}^{2}\right\} ^{1/2}.  \label{E03}
\end{equation}%
By replacing the first sign $\pm $ for $-$, we have expressions for the
lowest two levels, $E_{0,1}$. The difference between these energy levels at $%
\varepsilon =0$ is the minimal splitting:%
\begin{equation}
\Delta _{0}=\sqrt{\left[ \frac{E_{\mathrm{R}}}{2}+t_{\mathrm{d}}\right]
^{2}+t_{\mathrm{v}}^{2}}-\sqrt{\left[ \frac{E_{\mathrm{R}}}{2}-t_{\mathrm{d}}%
\right] ^{2}+t_{\mathrm{v}}^{2}}.  \label{Delta0}
\end{equation}%
Given this, for small $\varepsilon $ we can expand $\Delta E^{2}=\left(
E_{1}-E_{0}\right) ^{2}$ into series and obtain the spectrum%
\begin{equation}
E_{1}-E_{0}\approx \sqrt{\Delta _{0}^{2}+0.16\varepsilon ^{2}}.
\label{spectrum}
\end{equation}%
Note that this is similar to a qubit spectrum $\sqrt{\Delta
_{0}^{2}+\varepsilon ^{2}}$, but differs by a numerical factor. In Fig.~\ref%
{Fig2}(a) we can see that the qubit-like spectrum, Eq.~(\ref{spectrum}), is
sufficient for describing the dressed energy levels at small values of the
bias.

Symmetric heart-shaped resonances are shown in Fig.~\ref{Fig:heart}. Note
that with increasing the driving amplitude, the resonance lines change from
\textit{convex} to \textit{concave} shapes.

For the asymmetric case, with $E_{\mathrm{L}}\neq E_{\mathrm{R}}$, we can
generalize Eq.~(\ref{E03}) assuming
\begin{eqnarray}
E_{0/1} &=&\theta (\varepsilon )\left\{ \frac{E_{\mathrm{R}}}{2}-\left( %
\left[ \frac{E_{\mathrm{R}}}{2}\pm \sqrt{\left( \frac{\varepsilon }{2}%
\right) ^{2}+t_{\mathrm{d}}^{2}}\right] ^{2}+t_{\mathrm{v}}^{2}\right)
^{1/2}\right\}  \label{guess} \\
&&+\theta (-\varepsilon )\left\{ \frac{E_{\mathrm{L}}}{2}-\left( \left[
\frac{E_{\mathrm{L}}}{2}\pm \sqrt{\left( \frac{\varepsilon }{2}\right)
^{2}+t_{\mathrm{d}}^{2}}\right] ^{2}+t_{\mathrm{v}}^{2}\right)
^{1/2}\right\} .  \notag
\end{eqnarray}%
For large values of the bias, $\varepsilon \rightarrow \infty $, equation~(%
\ref{guess}) gives
\begin{equation}
E_{1}-E_{0}\approx \theta (\varepsilon )E_{\mathrm{R}}-\theta (-\varepsilon
)E_{\mathrm{L}},
\end{equation}%
which correctly describes the spectrum, demonstrated in Fig.~\ref{Fig2}(a).
Equation~(\ref{guess}) can be useful for analytical studies.

\nocite{apsrev41Control}
\bibliographystyle{apsrev4-1}
\bibliography{harp}

\begin{thebibliography}{45}%
\makeatletter
\providecommand \@ifxundefined [1]{%
 \@ifx{#1\undefined}
}%
\providecommand \@ifnum [1]{%
 \ifnum #1\expandafter \@firstoftwo
 \else \expandafter \@secondoftwo
 \fi
}%
\providecommand \@ifx [1]{%
 \ifx #1\expandafter \@firstoftwo
 \else \expandafter \@secondoftwo
 \fi
}%
\providecommand \natexlab [1]{#1}%
\providecommand \enquote  [1]{``#1''}%
\providecommand \bibnamefont  [1]{#1}%
\providecommand \bibfnamefont [1]{#1}%
\providecommand \citenamefont [1]{#1}%
\providecommand \href@noop [0]{\@secondoftwo}%
\providecommand \href [0]{\begingroup \@sanitize@url \@href}%
\providecommand \@href[1]{\@@startlink{#1}\@@href}%
\providecommand \@@href[1]{\endgroup#1\@@endlink}%
\providecommand \@sanitize@url [0]{\catcode `\\12\catcode `\$12\catcode
  `\&12\catcode `\#12\catcode `\^12\catcode `\_12\catcode `\%12\relax}%
\providecommand \@@startlink[1]{}%
\providecommand \@@endlink[0]{}%
\providecommand \url  [0]{\begingroup\@sanitize@url \@url }%
\providecommand \@url [1]{\endgroup\@href {#1}{\urlprefix }}%
\providecommand \urlprefix  [0]{URL }%
\providecommand \Eprint [0]{\href }%
\providecommand \doibase [0]{http://dx.doi.org/}%
\providecommand \selectlanguage [0]{\@gobble}%
\providecommand \bibinfo  [0]{\@secondoftwo}%
\providecommand \bibfield  [0]{\@secondoftwo}%
\providecommand \translation [1]{[#1]}%
\providecommand \BibitemOpen [0]{}%
\providecommand \bibitemStop [0]{}%
\providecommand \bibitemNoStop [0]{.\EOS\space}%
\providecommand \EOS [0]{\spacefactor3000\relax}%
\providecommand \BibitemShut  [1]{\csname bibitem#1\endcsname}%
\let\auto@bib@innerbib\@empty
\bibitem [{\citenamefont {Berry}(1995)}]{Berry95}%
  \BibitemOpen
  \bibfield  {author} {\bibinfo {author} {\bibfnamefont {M.}~\bibnamefont
  {Berry}},\ }\bibfield  {title} {\enquote {\bibinfo {title} {Two-state quantum
  asymptotics},}\ }\href@noop {} {\bibfield  {journal} {\bibinfo  {journal}
  {Annals of the New York Academy of Sciences}\ }\textbf {\bibinfo {volume}
  {755}},\ \bibinfo {pages} {303--317} (\bibinfo {year} {1995})}\BibitemShut
  {NoStop}%
\bibitem [{\citenamefont {Degen}\ \emph {et~al.}(2017)\citenamefont {Degen},
  \citenamefont {Reinhard},\ and\ \citenamefont {Cappellaro}}]{Degen17}%
  \BibitemOpen
  \bibfield  {author} {\bibinfo {author} {\bibfnamefont {C.~L.}\ \bibnamefont
  {Degen}}, \bibinfo {author} {\bibfnamefont {F.}~\bibnamefont {Reinhard}}, \
  and\ \bibinfo {author} {\bibfnamefont {P.}~\bibnamefont {Cappellaro}},\
  }\bibfield  {title} {\enquote {\bibinfo {title} {Quantum sensing},}\
  }\href@noop {} {\bibfield  {journal} {\bibinfo  {journal} {Rev. Mod. Phys.}\
  }\textbf {\bibinfo {volume} {89}},\ \bibinfo {pages} {035002} (\bibinfo
  {year} {2017})}\BibitemShut {NoStop}%
\bibitem [{\citenamefont {Buluta}\ \emph {et~al.}(2011)\citenamefont {Buluta},
  \citenamefont {Ashhab},\ and\ \citenamefont {Nori}}]{Buluta11}%
  \BibitemOpen
  \bibfield  {author} {\bibinfo {author} {\bibfnamefont {I.}~\bibnamefont
  {Buluta}}, \bibinfo {author} {\bibfnamefont {S.}~\bibnamefont {Ashhab}}, \
  and\ \bibinfo {author} {\bibfnamefont {F.}~\bibnamefont {Nori}},\ }\bibfield
  {title} {\enquote {\bibinfo {title} {Natural and artificial atoms for quantum
  computation},}\ }\href@noop {} {\bibfield  {journal} {\bibinfo  {journal}
  {Rep. Prog. Phys.}\ }\textbf {\bibinfo {volume} {74}},\ \bibinfo {pages}
  {104401} (\bibinfo {year} {2011})}\BibitemShut {NoStop}%
\bibitem [{\citenamefont {Shevchenko}\ \emph {et~al.}(2010)\citenamefont
  {Shevchenko}, \citenamefont {Ashhab},\ and\ \citenamefont
  {Nori}}]{Shevchenko10}%
  \BibitemOpen
  \bibfield  {author} {\bibinfo {author} {\bibfnamefont {S.~N.}\ \bibnamefont
  {Shevchenko}}, \bibinfo {author} {\bibfnamefont {S.}~\bibnamefont {Ashhab}},
  \ and\ \bibinfo {author} {\bibfnamefont {F.}~\bibnamefont {Nori}},\
  }\bibfield  {title} {\enquote {\bibinfo {title}
  {Landau-{Z}ener-{S}t{\"u}ckelberg interferometry},}\ }\href@noop {}
  {\bibfield  {journal} {\bibinfo  {journal} {Phys. Rep.}\ }\textbf {\bibinfo
  {volume} {492}},\ \bibinfo {pages} {1--30} (\bibinfo {year}
  {2010})}\BibitemShut {NoStop}%
\bibitem [{\citenamefont {Oliver}\ \emph {et~al.}(2005)\citenamefont {Oliver},
  \citenamefont {Yu}, \citenamefont {Lee}, \citenamefont {Berggren},
  \citenamefont {Levitov},\ and\ \citenamefont {Orlando}}]{Oliver05}%
  \BibitemOpen
  \bibfield  {author} {\bibinfo {author} {\bibfnamefont {W.~D.}\ \bibnamefont
  {Oliver}}, \bibinfo {author} {\bibfnamefont {Y.}~\bibnamefont {Yu}}, \bibinfo
  {author} {\bibfnamefont {J.~C.}\ \bibnamefont {Lee}}, \bibinfo {author}
  {\bibfnamefont {K.~K.}\ \bibnamefont {Berggren}}, \bibinfo {author}
  {\bibfnamefont {L.~S.}\ \bibnamefont {Levitov}}, \ and\ \bibinfo {author}
  {\bibfnamefont {T.~P.}\ \bibnamefont {Orlando}},\ }\bibfield  {title}
  {\enquote {\bibinfo {title} {{Mach-Zehnder} interferometry in a strongly
  driven superconducting qubit},}\ }\href@noop {} {\bibfield  {journal}
  {\bibinfo  {journal} {Science}\ }\textbf {\bibinfo {volume} {310}},\ \bibinfo
  {pages} {1653--1657} (\bibinfo {year} {2005})}\BibitemShut {NoStop}%
\bibitem [{\citenamefont {Sillanp\"a\"a}\ \emph {et~al.}(2006)\citenamefont
  {Sillanp\"a\"a}, \citenamefont {Lehtinen}, \citenamefont {Paila},
  \citenamefont {Makhlin},\ and\ \citenamefont {Hakonen}}]{Sillanpaa06}%
  \BibitemOpen
  \bibfield  {author} {\bibinfo {author} {\bibfnamefont {M.}~\bibnamefont
  {Sillanp\"a\"a}}, \bibinfo {author} {\bibfnamefont {T.}~\bibnamefont
  {Lehtinen}}, \bibinfo {author} {\bibfnamefont {A.}~\bibnamefont {Paila}},
  \bibinfo {author} {\bibfnamefont {Y.}~\bibnamefont {Makhlin}}, \ and\
  \bibinfo {author} {\bibfnamefont {P.}~\bibnamefont {Hakonen}},\ }\bibfield
  {title} {\enquote {\bibinfo {title} {Continuous-time monitoring of
  {Landau-Zener} interference in a {C}ooper-pair box},}\ }\href@noop {}
  {\bibfield  {journal} {\bibinfo  {journal} {Phys. Rev. Lett.}\ }\textbf
  {\bibinfo {volume} {96}},\ \bibinfo {pages} {187002} (\bibinfo {year}
  {2006})}\BibitemShut {NoStop}%
\bibitem [{\citenamefont {Wilson}\ \emph {et~al.}(2007)\citenamefont {Wilson},
  \citenamefont {Duty}, \citenamefont {Persson}, \citenamefont {Sandberg},
  \citenamefont {Johansson},\ and\ \citenamefont {Delsing}}]{Wilson07}%
  \BibitemOpen
  \bibfield  {author} {\bibinfo {author} {\bibfnamefont {C.~M.}\ \bibnamefont
  {Wilson}}, \bibinfo {author} {\bibfnamefont {T.}~\bibnamefont {Duty}},
  \bibinfo {author} {\bibfnamefont {F.}~\bibnamefont {Persson}}, \bibinfo
  {author} {\bibfnamefont {M.}~\bibnamefont {Sandberg}}, \bibinfo {author}
  {\bibfnamefont {G.}~\bibnamefont {Johansson}}, \ and\ \bibinfo {author}
  {\bibfnamefont {P.}~\bibnamefont {Delsing}},\ }\bibfield  {title} {\enquote
  {\bibinfo {title} {Coherence times of dressed states of a superconducting
  qubit under extreme driving},}\ }\href@noop {} {\bibfield  {journal}
  {\bibinfo  {journal} {Phys. Rev. Lett.}\ }\textbf {\bibinfo {volume} {98}},\
  \bibinfo {pages} {257003} (\bibinfo {year} {2007})}\BibitemShut {NoStop}%
\bibitem [{\citenamefont {Izmalkov}\ \emph {et~al.}(2008)\citenamefont
  {Izmalkov}, \citenamefont {van~der Ploeg}, \citenamefont {Shevchenko},
  \citenamefont {Grajcar}, \citenamefont {Il'ichev}, \citenamefont {H\"ubner},
  \citenamefont {Omelyanchouk},\ and\ \citenamefont {Meyer}}]{Izmalkov08}%
  \BibitemOpen
  \bibfield  {author} {\bibinfo {author} {\bibfnamefont {A.}~\bibnamefont
  {Izmalkov}}, \bibinfo {author} {\bibfnamefont {S.~H.~W.}\ \bibnamefont
  {van~der Ploeg}}, \bibinfo {author} {\bibfnamefont {S.~N.}\ \bibnamefont
  {Shevchenko}}, \bibinfo {author} {\bibfnamefont {M.}~\bibnamefont {Grajcar}},
  \bibinfo {author} {\bibfnamefont {E.}~\bibnamefont {Il'ichev}}, \bibinfo
  {author} {\bibfnamefont {U.}~\bibnamefont {H\"ubner}}, \bibinfo {author}
  {\bibfnamefont {A.~N.}\ \bibnamefont {Omelyanchouk}}, \ and\ \bibinfo
  {author} {\bibfnamefont {H.-G.}\ \bibnamefont {Meyer}},\ }\bibfield  {title}
  {\enquote {\bibinfo {title} {Consistency of ground state and spectroscopic
  measurements on flux qubits},}\ }\href@noop {} {\bibfield  {journal}
  {\bibinfo  {journal} {Phys. Rev. Lett.}\ }\textbf {\bibinfo {volume} {101}},\
  \bibinfo {pages} {017003} (\bibinfo {year} {2008})}\BibitemShut {NoStop}%
\bibitem [{\citenamefont {Sun}\ \emph {et~al.}(2009)\citenamefont {Sun},
  \citenamefont {Wen}, \citenamefont {Wang}, \citenamefont {Cong},
  \citenamefont {Chen}, \citenamefont {Kang}, \citenamefont {Xu}, \citenamefont
  {Yu}, \citenamefont {Han},\ and\ \citenamefont {Wu}}]{Sun09}%
  \BibitemOpen
  \bibfield  {author} {\bibinfo {author} {\bibfnamefont {G.}~\bibnamefont
  {Sun}}, \bibinfo {author} {\bibfnamefont {X.}~\bibnamefont {Wen}}, \bibinfo
  {author} {\bibfnamefont {Y.}~\bibnamefont {Wang}}, \bibinfo {author}
  {\bibfnamefont {S.}~\bibnamefont {Cong}}, \bibinfo {author} {\bibfnamefont
  {J.}~\bibnamefont {Chen}}, \bibinfo {author} {\bibfnamefont {L.}~\bibnamefont
  {Kang}}, \bibinfo {author} {\bibfnamefont {W.}~\bibnamefont {Xu}}, \bibinfo
  {author} {\bibfnamefont {Y.}~\bibnamefont {Yu}}, \bibinfo {author}
  {\bibfnamefont {S.}~\bibnamefont {Han}}, \ and\ \bibinfo {author}
  {\bibfnamefont {P.}~\bibnamefont {Wu}},\ }\bibfield  {title} {\enquote
  {\bibinfo {title} {Population inversion induced by {Landau-Zener} transition
  in a strongly driven rf superconducting quantum interference device},}\
  }\href@noop {} {\bibfield  {journal} {\bibinfo  {journal} {Appl. Phys.
  Lett.}\ }\textbf {\bibinfo {volume} {94}},\ \bibinfo {pages} {102502}
  (\bibinfo {year} {2009})}\BibitemShut {NoStop}%
\bibitem [{\citenamefont {Stehlik}\ \emph {et~al.}(2012)\citenamefont
  {Stehlik}, \citenamefont {Dovzhenko}, \citenamefont {Petta}, \citenamefont
  {Johansson}, \citenamefont {Nori}, \citenamefont {Lu},\ and\ \citenamefont
  {Gossard}}]{Stehlik12}%
  \BibitemOpen
  \bibfield  {author} {\bibinfo {author} {\bibfnamefont {J.}~\bibnamefont
  {Stehlik}}, \bibinfo {author} {\bibfnamefont {Y.}~\bibnamefont {Dovzhenko}},
  \bibinfo {author} {\bibfnamefont {J.~R.}\ \bibnamefont {Petta}}, \bibinfo
  {author} {\bibfnamefont {J.~R.}\ \bibnamefont {Johansson}}, \bibinfo {author}
  {\bibfnamefont {F.}~\bibnamefont {Nori}}, \bibinfo {author} {\bibfnamefont
  {H.}~\bibnamefont {Lu}}, \ and\ \bibinfo {author} {\bibfnamefont {A.~C.}\
  \bibnamefont {Gossard}},\ }\bibfield  {title} {\enquote {\bibinfo {title}
  {{Landau-Zener-St\"uckelberg} interferometry of a single electron charge
  qubit},}\ }\href@noop {} {\bibfield  {journal} {\bibinfo  {journal} {Phys.
  Rev. B}\ }\textbf {\bibinfo {volume} {86}},\ \bibinfo {pages} {121303}
  (\bibinfo {year} {2012})}\BibitemShut {NoStop}%
\bibitem [{\citenamefont {Gonzalez-Zalba}\ \emph {et~al.}(2016)\citenamefont
  {Gonzalez-Zalba}, \citenamefont {Shevchenko}, \citenamefont {Barraud},
  \citenamefont {Johansson}, \citenamefont {Ferguson}, \citenamefont {Nori},\
  and\ \citenamefont {Betz}}]{Gonzalez-Zalba16}%
  \BibitemOpen
  \bibfield  {author} {\bibinfo {author} {\bibfnamefont {M.~F.}\ \bibnamefont
  {Gonzalez-Zalba}}, \bibinfo {author} {\bibfnamefont {S.~N.}\ \bibnamefont
  {Shevchenko}}, \bibinfo {author} {\bibfnamefont {S.}~\bibnamefont {Barraud}},
  \bibinfo {author} {\bibfnamefont {J.~R.}\ \bibnamefont {Johansson}}, \bibinfo
  {author} {\bibfnamefont {A.~J.}\ \bibnamefont {Ferguson}}, \bibinfo {author}
  {\bibfnamefont {F.}~\bibnamefont {Nori}}, \ and\ \bibinfo {author}
  {\bibfnamefont {A.~C.}\ \bibnamefont {Betz}},\ }\bibfield  {title} {\enquote
  {\bibinfo {title} {Gate-sensing coherent charge oscillations in a silicon
  field-effect transistor},}\ }\href@noop {} {\bibfield  {journal} {\bibinfo
  {journal} {Nano Lett.}\ }\textbf {\bibinfo {volume} {16}},\ \bibinfo {pages}
  {1614--1619} (\bibinfo {year} {2016})}\BibitemShut {NoStop}%
\bibitem [{\citenamefont {Berns}\ \emph {et~al.}(2008)\citenamefont {Berns},
  \citenamefont {Rudner}, \citenamefont {Valenzuela}, \citenamefont {Berggren},
  \citenamefont {Oliver}, \citenamefont {Levitov},\ and\ \citenamefont
  {Orlando}}]{Berns08}%
  \BibitemOpen
  \bibfield  {author} {\bibinfo {author} {\bibfnamefont {D.~M.}\ \bibnamefont
  {Berns}}, \bibinfo {author} {\bibfnamefont {M.~S.}\ \bibnamefont {Rudner}},
  \bibinfo {author} {\bibfnamefont {S.~O.}\ \bibnamefont {Valenzuela}},
  \bibinfo {author} {\bibfnamefont {K.~K.}\ \bibnamefont {Berggren}}, \bibinfo
  {author} {\bibfnamefont {W.~D.}\ \bibnamefont {Oliver}}, \bibinfo {author}
  {\bibfnamefont {L.~S.}\ \bibnamefont {Levitov}}, \ and\ \bibinfo {author}
  {\bibfnamefont {T.~P.}\ \bibnamefont {Orlando}},\ }\bibfield  {title}
  {\enquote {\bibinfo {title} {Amplitude spectroscopy of a solid-state
  artificial atom},}\ }\href@noop {} {\bibfield  {journal} {\bibinfo  {journal}
  {Nature}\ }\textbf {\bibinfo {volume} {455}},\ \bibinfo {pages} {51}
  (\bibinfo {year} {2008})}\BibitemShut {NoStop}%
\bibitem [{\citenamefont {Satanin}\ \emph {et~al.}(2012)\citenamefont
  {Satanin}, \citenamefont {Denisenko}, \citenamefont {Ashhab},\ and\
  \citenamefont {Nori}}]{Satanin12}%
  \BibitemOpen
  \bibfield  {author} {\bibinfo {author} {\bibfnamefont {A.~M.}\ \bibnamefont
  {Satanin}}, \bibinfo {author} {\bibfnamefont {M.~V.}\ \bibnamefont
  {Denisenko}}, \bibinfo {author} {\bibfnamefont {S.}~\bibnamefont {Ashhab}}, \
  and\ \bibinfo {author} {\bibfnamefont {F.}~\bibnamefont {Nori}},\ }\bibfield
  {title} {\enquote {\bibinfo {title} {Amplitude spectroscopy of two coupled
  qubits},}\ }\href@noop {} {\bibfield  {journal} {\bibinfo  {journal} {Phys.
  Rev. B}\ }\textbf {\bibinfo {volume} {85}},\ \bibinfo {pages} {184524}
  (\bibinfo {year} {2012})}\BibitemShut {NoStop}%
\bibitem [{\citenamefont {Sun}\ \emph {et~al.}(2011)\citenamefont {Sun},
  \citenamefont {Wen}, \citenamefont {Mao}, \citenamefont {Yu}, \citenamefont
  {Chen}, \citenamefont {Xu}, \citenamefont {Kang}, \citenamefont {Wu},\ and\
  \citenamefont {Han}}]{Sun11}%
  \BibitemOpen
  \bibfield  {author} {\bibinfo {author} {\bibfnamefont {G.}~\bibnamefont
  {Sun}}, \bibinfo {author} {\bibfnamefont {X.}~\bibnamefont {Wen}}, \bibinfo
  {author} {\bibfnamefont {B.}~\bibnamefont {Mao}}, \bibinfo {author}
  {\bibfnamefont {Y.}~\bibnamefont {Yu}}, \bibinfo {author} {\bibfnamefont
  {J.}~\bibnamefont {Chen}}, \bibinfo {author} {\bibfnamefont {W.}~\bibnamefont
  {Xu}}, \bibinfo {author} {\bibfnamefont {L.}~\bibnamefont {Kang}}, \bibinfo
  {author} {\bibfnamefont {P.}~\bibnamefont {Wu}}, \ and\ \bibinfo {author}
  {\bibfnamefont {S.}~\bibnamefont {Han}},\ }\bibfield  {title} {\enquote
  {\bibinfo {title} {{Landau-Zener-St\"uckelberg} interference of
  microwave-dressed states of a superconducting phase qubit},}\ }\href@noop {}
  {\bibfield  {journal} {\bibinfo  {journal} {Phys. Rev. B}\ }\textbf {\bibinfo
  {volume} {83}},\ \bibinfo {pages} {180507} (\bibinfo {year}
  {2011})}\BibitemShut {NoStop}%
\bibitem [{\citenamefont {Gong}\ \emph {et~al.}(2016)\citenamefont {Gong},
  \citenamefont {Zhou}, \citenamefont {Lan}, \citenamefont {Fan}, \citenamefont
  {Pan}, \citenamefont {Yu}, \citenamefont {Chen}, \citenamefont {Sun},
  \citenamefont {Yu}, \citenamefont {Han},\ and\ \citenamefont {Wu}}]{Gong16}%
  \BibitemOpen
  \bibfield  {author} {\bibinfo {author} {\bibfnamefont {M.}~\bibnamefont
  {Gong}}, \bibinfo {author} {\bibfnamefont {Y.}~\bibnamefont {Zhou}}, \bibinfo
  {author} {\bibfnamefont {D.}~\bibnamefont {Lan}}, \bibinfo {author}
  {\bibfnamefont {Y.}~\bibnamefont {Fan}}, \bibinfo {author} {\bibfnamefont
  {J.}~\bibnamefont {Pan}}, \bibinfo {author} {\bibfnamefont {H.}~\bibnamefont
  {Yu}}, \bibinfo {author} {\bibfnamefont {J.}~\bibnamefont {Chen}}, \bibinfo
  {author} {\bibfnamefont {G.}~\bibnamefont {Sun}}, \bibinfo {author}
  {\bibfnamefont {Y.}~\bibnamefont {Yu}}, \bibinfo {author} {\bibfnamefont
  {S.}~\bibnamefont {Han}}, \ and\ \bibinfo {author} {\bibfnamefont
  {P.}~\bibnamefont {Wu}},\ }\bibfield  {title} {\enquote {\bibinfo {title}
  {{Landau-Zener-St\"uckelberg-Majorana} interference in a {3D} transmon driven
  by a chirped microwave},}\ }\href@noop {} {\bibfield  {journal} {\bibinfo
  {journal} {Appl. Phys. Lett.}\ }\textbf {\bibinfo {volume} {108}},\ \bibinfo
  {eid} {112602} (\bibinfo {year} {2016})}\BibitemShut {NoStop}%
\bibitem [{\citenamefont {Greenberg}(2007)}]{Greenberg07}%
  \BibitemOpen
  \bibfield  {author} {\bibinfo {author} {\bibfnamefont {Y.~S.}\ \bibnamefont
  {Greenberg}},\ }\bibfield  {title} {\enquote {\bibinfo {title} {Low-frequency
  {R}abi spectroscopy of dissipative two-level systems: Dressed-state
  approach},}\ }\href@noop {} {\bibfield  {journal} {\bibinfo  {journal} {Phys.
  Rev. B}\ }\textbf {\bibinfo {volume} {76}},\ \bibinfo {pages} {104520}
  (\bibinfo {year} {2007})}\BibitemShut {NoStop}%
\bibitem [{\citenamefont {Greenberg}\ and\ \citenamefont
  {Il'ichev}(2008)}]{Greenberg08}%
  \BibitemOpen
  \bibfield  {author} {\bibinfo {author} {\bibfnamefont {Y.~S.}\ \bibnamefont
  {Greenberg}}\ and\ \bibinfo {author} {\bibfnamefont {E.}~\bibnamefont
  {Il'ichev}},\ }\bibfield  {title} {\enquote {\bibinfo {title} {Quantum theory
  of the low-frequency linear susceptibility of interferometer-type
  superconducting qubits},}\ }\href@noop {} {\bibfield  {journal} {\bibinfo
  {journal} {Phys. Rev. B}\ }\textbf {\bibinfo {volume} {77}},\ \bibinfo
  {pages} {094513} (\bibinfo {year} {2008})}\BibitemShut {NoStop}%
\bibitem [{\citenamefont {Mefed}(1999)}]{Mefed99}%
  \BibitemOpen
  \bibfield  {author} {\bibinfo {author} {\bibfnamefont {A.~E.}\ \bibnamefont
  {Mefed}},\ }\bibfield  {title} {\enquote {\bibinfo {title} {Spectrometer for
  studying {NMR} and relaxation in the doubly rotating frame},}\ }\href@noop {}
  {\bibfield  {journal} {\bibinfo  {journal} {Applied Magnetic Resonance}\
  }\textbf {\bibinfo {volume} {16}},\ \bibinfo {pages} {411--426} (\bibinfo
  {year} {1999})}\BibitemShut {NoStop}%
\bibitem [{\citenamefont {Tuorila}\ \emph {et~al.}(2010)\citenamefont
  {Tuorila}, \citenamefont {Silveri}, \citenamefont {Sillanp\"a\"a},
  \citenamefont {Thuneberg}, \citenamefont {Makhlin},\ and\ \citenamefont
  {Hakonen}}]{Tuorila10}%
  \BibitemOpen
  \bibfield  {author} {\bibinfo {author} {\bibfnamefont {J.}~\bibnamefont
  {Tuorila}}, \bibinfo {author} {\bibfnamefont {M.}~\bibnamefont {Silveri}},
  \bibinfo {author} {\bibfnamefont {M.}~\bibnamefont {Sillanp\"a\"a}}, \bibinfo
  {author} {\bibfnamefont {E.}~\bibnamefont {Thuneberg}}, \bibinfo {author}
  {\bibfnamefont {Y.}~\bibnamefont {Makhlin}}, \ and\ \bibinfo {author}
  {\bibfnamefont {P.}~\bibnamefont {Hakonen}},\ }\bibfield  {title} {\enquote
  {\bibinfo {title} {Stark effect and generalized {Bloch-Siegert} shift in a
  strongly driven two-level system},}\ }\href@noop {} {\bibfield  {journal}
  {\bibinfo  {journal} {Phys. Rev. Lett.}\ }\textbf {\bibinfo {volume} {105}},\
  \bibinfo {pages} {257003} (\bibinfo {year} {2010})}\BibitemShut {NoStop}%
\bibitem [{\citenamefont {Silveri}\ \emph {et~al.}(2013)\citenamefont
  {Silveri}, \citenamefont {Tuorila}, \citenamefont {Kemppainen},\ and\
  \citenamefont {Thuneberg}}]{Silveri13}%
  \BibitemOpen
  \bibfield  {author} {\bibinfo {author} {\bibfnamefont {M.}~\bibnamefont
  {Silveri}}, \bibinfo {author} {\bibfnamefont {J.}~\bibnamefont {Tuorila}},
  \bibinfo {author} {\bibfnamefont {M.}~\bibnamefont {Kemppainen}}, \ and\
  \bibinfo {author} {\bibfnamefont {E.}~\bibnamefont {Thuneberg}},\ }\bibfield
  {title} {\enquote {\bibinfo {title} {Probe spectroscopy of quasienergy
  states},}\ }\href@noop {} {\bibfield  {journal} {\bibinfo  {journal} {Phys.
  Rev. B}\ }\textbf {\bibinfo {volume} {87}},\ \bibinfo {pages} {134505}
  (\bibinfo {year} {2013})}\BibitemShut {NoStop}%
\bibitem [{\citenamefont {Saiko}\ \emph {et~al.}(2014)\citenamefont {Saiko},
  \citenamefont {Fedaruk},\ and\ \citenamefont {Markevich}}]{Saiko14}%
  \BibitemOpen
  \bibfield  {author} {\bibinfo {author} {\bibfnamefont {A.~P.}\ \bibnamefont
  {Saiko}}, \bibinfo {author} {\bibfnamefont {R.}~\bibnamefont {Fedaruk}}, \
  and\ \bibinfo {author} {\bibfnamefont {S.~A.}\ \bibnamefont {Markevich}},\
  }\bibfield  {title} {\enquote {\bibinfo {title} {Relaxation, decoherence, and
  steady-state population inversion in qubits doubly dressed by microwave and
  radiofrequency fields},}\ }\href@noop {} {\bibfield  {journal} {\bibinfo
  {journal} {J. Phys. B}\ }\textbf {\bibinfo {volume} {47}},\ \bibinfo {pages}
  {155502} (\bibinfo {year} {2014})}\BibitemShut {NoStop}%
\bibitem [{\citenamefont {Neilinger}\ \emph {et~al.}(2016)\citenamefont
  {Neilinger}, \citenamefont {Shevchenko}, \citenamefont {Bog\'ar},
  \citenamefont {Reh\'ak}, \citenamefont {Oelsner}, \citenamefont {Karpov},
  \citenamefont {H\"ubner}, \citenamefont {Astafiev}, \citenamefont {Grajcar},\
  and\ \citenamefont {Il'ichev}}]{Neilinger16}%
  \BibitemOpen
  \bibfield  {author} {\bibinfo {author} {\bibfnamefont {P.}~\bibnamefont
  {Neilinger}}, \bibinfo {author} {\bibfnamefont {S.~N.}\ \bibnamefont
  {Shevchenko}}, \bibinfo {author} {\bibfnamefont {J.}~\bibnamefont {Bog\'ar}},
  \bibinfo {author} {\bibfnamefont {M.}~\bibnamefont {Reh\'ak}}, \bibinfo
  {author} {\bibfnamefont {G.}~\bibnamefont {Oelsner}}, \bibinfo {author}
  {\bibfnamefont {D.~S.}\ \bibnamefont {Karpov}}, \bibinfo {author}
  {\bibfnamefont {U.}~\bibnamefont {H\"ubner}}, \bibinfo {author}
  {\bibfnamefont {O.}~\bibnamefont {Astafiev}}, \bibinfo {author}
  {\bibfnamefont {M.}~\bibnamefont {Grajcar}}, \ and\ \bibinfo {author}
  {\bibfnamefont {E.}~\bibnamefont {Il'ichev}},\ }\bibfield  {title} {\enquote
  {\bibinfo {title} {{Landau-Zener-St\"uckelberg-Majorana} lasing in circuit
  quantum electrodynamics},}\ }\href@noop {} {\bibfield  {journal} {\bibinfo
  {journal} {Phys. Rev. B}\ }\textbf {\bibinfo {volume} {94}},\ \bibinfo
  {pages} {094519} (\bibinfo {year} {2016})}\BibitemShut {NoStop}%
\bibitem [{\citenamefont {Jin-Dan}\ \emph {et~al.}(2011)\citenamefont
  {Jin-Dan}, \citenamefont {Xue-Da}, \citenamefont {Guo-Zhu},\ and\
  \citenamefont {Yang}}]{Chen11}%
  \BibitemOpen
  \bibfield  {author} {\bibinfo {author} {\bibfnamefont {C.}~\bibnamefont
  {Jin-Dan}}, \bibinfo {author} {\bibfnamefont {W.}~\bibnamefont {Xue-Da}},
  \bibinfo {author} {\bibfnamefont {S.}~\bibnamefont {Guo-Zhu}}, \ and\
  \bibinfo {author} {\bibfnamefont {Y.}~\bibnamefont {Yang}},\ }\bibfield
  {title} {\enquote {\bibinfo {title} {{Landau-Zener-St\"uckelberg}
  interference in a multi-anticrossing system},}\ }\href@noop {} {\bibfield
  {journal} {\bibinfo  {journal} {Chinese Phys. B}\ }\textbf {\bibinfo {volume}
  {20}},\ \bibinfo {pages} {088501} (\bibinfo {year} {2011})}\BibitemShut
  {NoStop}%
\bibitem [{\citenamefont {Kenmoe}\ \emph {et~al.}(2013)\citenamefont {Kenmoe},
  \citenamefont {Phien}, \citenamefont {Kiselev},\ and\ \citenamefont
  {Fai}}]{Kenmoe13}%
  \BibitemOpen
  \bibfield  {author} {\bibinfo {author} {\bibfnamefont {M.~B.}\ \bibnamefont
  {Kenmoe}}, \bibinfo {author} {\bibfnamefont {H.~N.}\ \bibnamefont {Phien}},
  \bibinfo {author} {\bibfnamefont {M.~N.}\ \bibnamefont {Kiselev}}, \ and\
  \bibinfo {author} {\bibfnamefont {L.~C.}\ \bibnamefont {Fai}},\ }\bibfield
  {title} {\enquote {\bibinfo {title} {Effects of colored noise on
  {Landau-Zener} transitions: Two- and three-level systems},}\ }\href@noop {}
  {\bibfield  {journal} {\bibinfo  {journal} {Phys. Rev. B}\ }\textbf {\bibinfo
  {volume} {87}},\ \bibinfo {pages} {224301} (\bibinfo {year}
  {2013})}\BibitemShut {NoStop}%
\bibitem [{\citenamefont {Ashhab}(2016)}]{Ashhab16}%
  \BibitemOpen
  \bibfield  {author} {\bibinfo {author} {\bibfnamefont {S.}~\bibnamefont
  {Ashhab}},\ }\bibfield  {title} {\enquote {\bibinfo {title} {{Landau-Zener}
  transitions in an open multilevel quantum system},}\ }\href@noop {}
  {\bibfield  {journal} {\bibinfo  {journal} {Phys. Rev. A}\ }\textbf {\bibinfo
  {volume} {94}},\ \bibinfo {pages} {042109} (\bibinfo {year}
  {2016})}\BibitemShut {NoStop}%
\bibitem [{\citenamefont {Stehlik}\ \emph {et~al.}(2016)\citenamefont
  {Stehlik}, \citenamefont {Maialle}, \citenamefont {Degani},\ and\
  \citenamefont {Petta}}]{Stehlik16}%
  \BibitemOpen
  \bibfield  {author} {\bibinfo {author} {\bibfnamefont {J.}~\bibnamefont
  {Stehlik}}, \bibinfo {author} {\bibfnamefont {M.~Z.}\ \bibnamefont
  {Maialle}}, \bibinfo {author} {\bibfnamefont {M.~H.}\ \bibnamefont {Degani}},
  \ and\ \bibinfo {author} {\bibfnamefont {J.~R.}\ \bibnamefont {Petta}},\
  }\bibfield  {title} {\enquote {\bibinfo {title} {Role of multilevel
  {Landau-Zener} interference in extreme harmonic generation},}\ }\href@noop {}
  {\bibfield  {journal} {\bibinfo  {journal} {Phys. Rev. B}\ }\textbf {\bibinfo
  {volume} {94}},\ \bibinfo {pages} {075307} (\bibinfo {year}
  {2016})}\BibitemShut {NoStop}%
\bibitem [{\citenamefont {Sinitsyn}\ and\ \citenamefont
  {Chernyak}(2017)}]{Sinitsyn17}%
  \BibitemOpen
  \bibfield  {author} {\bibinfo {author} {\bibfnamefont {N.~A.}\ \bibnamefont
  {Sinitsyn}}\ and\ \bibinfo {author} {\bibfnamefont {V.~Y.}\ \bibnamefont
  {Chernyak}},\ }\bibfield  {title} {\enquote {\bibinfo {title} {The quest for
  solvable multistate {Landau-Zener} models},}\ }\href@noop {} {\bibfield
  {journal} {\bibinfo  {journal} {J.Phys. A: Math. Theor.}\ }\textbf {\bibinfo
  {volume} {50}},\ \bibinfo {pages} {255203} (\bibinfo {year}
  {2017})}\BibitemShut {NoStop}%
\bibitem [{\citenamefont {Chatterjee}\ \emph {et~al.}(2018)\citenamefont
  {Chatterjee}, \citenamefont {Shevchenko}, \citenamefont {Barraud},
  \citenamefont {Otxoa}, \citenamefont {Nori}, \citenamefont {Morton},\ and\
  \citenamefont {Gonzalez-Zalba}}]{Chatterjee18}%
  \BibitemOpen
  \bibfield  {author} {\bibinfo {author} {\bibfnamefont {A.}~\bibnamefont
  {Chatterjee}}, \bibinfo {author} {\bibfnamefont {S.~N.}\ \bibnamefont
  {Shevchenko}}, \bibinfo {author} {\bibfnamefont {S.}~\bibnamefont {Barraud}},
  \bibinfo {author} {\bibfnamefont {R.~M.}\ \bibnamefont {Otxoa}}, \bibinfo
  {author} {\bibfnamefont {F.}~\bibnamefont {Nori}}, \bibinfo {author}
  {\bibfnamefont {J.~J.~L.}\ \bibnamefont {Morton}}, \ and\ \bibinfo {author}
  {\bibfnamefont {M.~F.}\ \bibnamefont {Gonzalez-Zalba}},\ }\bibfield  {title}
  {\enquote {\bibinfo {title} {A silicon-based single-electron interferometer
  coupled to a fermionic sea},}\ }\href@noop {} {\bibfield  {journal} {\bibinfo
   {journal} {Phys. Rev. B}\ }\textbf {\bibinfo {volume} {97}},\ \bibinfo
  {pages} {045405} (\bibinfo {year} {2018})}\BibitemShut {NoStop}%
\bibitem [{\citenamefont {Bogan}\ \emph {et~al.}(2018)\citenamefont {Bogan},
  \citenamefont {Studenikin}, \citenamefont {Korkusinski}, \citenamefont
  {Gaudreau}, \citenamefont {Zawadzki}, \citenamefont {Sachrajda},
  \citenamefont {Tracy}, \citenamefont {Reno},\ and\ \citenamefont
  {Hargett}}]{Bogan18}%
  \BibitemOpen
  \bibfield  {author} {\bibinfo {author} {\bibfnamefont {A.}~\bibnamefont
  {Bogan}}, \bibinfo {author} {\bibfnamefont {S.}~\bibnamefont {Studenikin}},
  \bibinfo {author} {\bibfnamefont {M.}~\bibnamefont {Korkusinski}}, \bibinfo
  {author} {\bibfnamefont {L.}~\bibnamefont {Gaudreau}}, \bibinfo {author}
  {\bibfnamefont {P.}~\bibnamefont {Zawadzki}}, \bibinfo {author}
  {\bibfnamefont {A.~S.}\ \bibnamefont {Sachrajda}}, \bibinfo {author}
  {\bibfnamefont {L.}~\bibnamefont {Tracy}}, \bibinfo {author} {\bibfnamefont
  {J.}~\bibnamefont {Reno}}, \ and\ \bibinfo {author} {\bibfnamefont
  {T.}~\bibnamefont {Hargett}},\ }\bibfield  {title} {\enquote {\bibinfo
  {title} {{Landau-Zener-St\"uckelberg-Majorana} interferometry of a single
  hole},}\ }\href@noop {} {\bibfield  {journal} {\bibinfo  {journal} {Phys.
  Rev. Lett.}\ }\textbf {\bibinfo {volume} {120}},\ \bibinfo {pages} {207701}
  (\bibinfo {year} {2018})}\BibitemShut {NoStop}%
\bibitem [{\citenamefont {Koski}\ \emph {et~al.}(2018)\citenamefont {Koski},
  \citenamefont {Landig}, \citenamefont {Palyi}, \citenamefont {Scarlino},
  \citenamefont {Reichl}, \citenamefont {Wegscheider}, \citenamefont {Burkard},
  \citenamefont {Wallraff}, \citenamefont {Ensslin},\ and\ \citenamefont
  {Ihn}}]{Koski18}%
  \BibitemOpen
  \bibfield  {author} {\bibinfo {author} {\bibfnamefont {J.~V.}\ \bibnamefont
  {Koski}}, \bibinfo {author} {\bibfnamefont {A.~J.}\ \bibnamefont {Landig}},
  \bibinfo {author} {\bibfnamefont {A.}~\bibnamefont {Palyi}}, \bibinfo
  {author} {\bibfnamefont {P.}~\bibnamefont {Scarlino}}, \bibinfo {author}
  {\bibfnamefont {C.}~\bibnamefont {Reichl}}, \bibinfo {author} {\bibfnamefont
  {W.}~\bibnamefont {Wegscheider}}, \bibinfo {author} {\bibfnamefont
  {G.}~\bibnamefont {Burkard}}, \bibinfo {author} {\bibfnamefont
  {A.}~\bibnamefont {Wallraff}}, \bibinfo {author} {\bibfnamefont
  {K.}~\bibnamefont {Ensslin}}, \ and\ \bibinfo {author} {\bibfnamefont
  {T.}~\bibnamefont {Ihn}},\ }\bibfield  {title} {\enquote {\bibinfo {title}
  {Floquet spectroscopy of a strongly driven quantum dot charge qubit with a
  microwave resonator},}\ }\href@noop {} {\bibfield  {journal} {\bibinfo
  {journal} {Phys. Rev. Lett.}\ }\textbf {\bibinfo {volume} {121}},\ \bibinfo
  {pages} {043603} (\bibinfo {year} {2018})}\BibitemShut {NoStop}%
\bibitem [{\citenamefont {Gramajo}\ \emph {et~al.}(2018)\citenamefont
  {Gramajo}, \citenamefont {Dominguez},\ and\ \citenamefont
  {Sanchez}}]{Gramajo18}%
  \BibitemOpen
  \bibfield  {author} {\bibinfo {author} {\bibfnamefont {A.~L.}\ \bibnamefont
  {Gramajo}}, \bibinfo {author} {\bibfnamefont {D.}~\bibnamefont {Dominguez}},
  \ and\ \bibinfo {author} {\bibfnamefont {M.~J.}\ \bibnamefont {Sanchez}},\
  }\bibfield  {title} {\enquote {\bibinfo {title} {Amplitude tuning of steady
  state entanglement in strongly driven coupled qubits},}\ }\href@noop {}
  {\bibfield  {journal} {\bibinfo  {journal} {Phys. Rev. A}\ }\textbf {\bibinfo
  {volume} {98}},\ \bibinfo {pages} {042337} (\bibinfo {year}
  {2018})}\BibitemShut {NoStop}%
\bibitem [{\citenamefont {Parafilo}\ and\ \citenamefont
  {Kiselev}()}]{Parafilo18}%
  \BibitemOpen
  \bibfield  {author} {\bibinfo {author} {\bibfnamefont {A.~V.}\ \bibnamefont
  {Parafilo}}\ and\ \bibinfo {author} {\bibfnamefont {M.~N.}\ \bibnamefont
  {Kiselev}},\ }\bibfield  {title} {\enquote {\bibinfo {title} {{Landau-Zener}
  transitions and {R}abi oscillations in a {C}ooper-pair box: Beyond two-level
  models},}\ }\href@noop {} {\bibinfo  {journal} {arXiv:1807.11604}\
  }\BibitemShut {NoStop}%
\bibitem [{\citenamefont {Yang}\ \emph {et~al.}(2013)\citenamefont {Yang},
  \citenamefont {Rossi}, \citenamefont {Ruskov}, \citenamefont {Lai},
  \citenamefont {Mohiyaddin}, \citenamefont {Lee}, \citenamefont {Tahan},
  \citenamefont {Klimeck}, \citenamefont {Morello},\ and\ \citenamefont
  {Dzurak}}]{Yang13}%
  \BibitemOpen
\bibfield  {journal} {  }\bibfield  {author} {\bibinfo {author} {\bibfnamefont
  {C.~H.}\ \bibnamefont {Yang}}, \bibinfo {author} {\bibfnamefont
  {A.}~\bibnamefont {Rossi}}, \bibinfo {author} {\bibfnamefont
  {R.}~\bibnamefont {Ruskov}}, \bibinfo {author} {\bibfnamefont {N.~S.}\
  \bibnamefont {Lai}}, \bibinfo {author} {\bibfnamefont {F.~A.}\ \bibnamefont
  {Mohiyaddin}}, \bibinfo {author} {\bibfnamefont {S.}~\bibnamefont {Lee}},
  \bibinfo {author} {\bibfnamefont {C.}~\bibnamefont {Tahan}}, \bibinfo
  {author} {\bibfnamefont {G.}~\bibnamefont {Klimeck}}, \bibinfo {author}
  {\bibfnamefont {A.}~\bibnamefont {Morello}}, \ and\ \bibinfo {author}
  {\bibfnamefont {A.~S.}\ \bibnamefont {Dzurak}},\ }\bibfield  {title}
  {\enquote {\bibinfo {title} {Spin-valley lifetimes in a silicon quantum dot
  with tunable valley splitting},}\ }\href@noop {} {\bibfield  {journal}
  {\bibinfo  {journal} {Nature Comm.}\ }\textbf {\bibinfo {volume} {4}},\
  \bibinfo {pages} {2069} (\bibinfo {year} {2013})}\BibitemShut {NoStop}%
\bibitem [{\citenamefont {Burkard}\ and\ \citenamefont
  {Petta}(2016)}]{Burkard16}%
  \BibitemOpen
  \bibfield  {author} {\bibinfo {author} {\bibfnamefont {G.}~\bibnamefont
  {Burkard}}\ and\ \bibinfo {author} {\bibfnamefont {J.~R.}\ \bibnamefont
  {Petta}},\ }\bibfield  {title} {\enquote {\bibinfo {title} {Dispersive
  readout of valley splittings in cavity-coupled silicon quantum dots},}\
  }\href@noop {} {\bibfield  {journal} {\bibinfo  {journal} {Phys. Rev. B}\
  }\textbf {\bibinfo {volume} {94}},\ \bibinfo {pages} {195305} (\bibinfo
  {year} {2016})}\BibitemShut {NoStop}%
\bibitem [{\citenamefont {Zhao}\ and\ \citenamefont {Hu}(2018)}]{Zhao18}%
  \BibitemOpen
  \bibfield  {author} {\bibinfo {author} {\bibfnamefont {X.}~\bibnamefont
  {Zhao}}\ and\ \bibinfo {author} {\bibfnamefont {X.}~\bibnamefont {Hu}},\
  }\bibfield  {title} {\enquote {\bibinfo {title} {Coherent electron transport
  in silicon quantum dots},}\ }\href@noop {} {\bibfield  {journal} {\bibinfo
  {journal} {arXiv:1803.00749}\ } (\bibinfo {year} {2018})}\BibitemShut
  {NoStop}%
\bibitem [{\citenamefont {Mi}\ \emph {et~al.}(2018)\citenamefont {Mi},
  \citenamefont {Kohler},\ and\ \citenamefont {Petta}}]{Mi18}%
  \BibitemOpen
  \bibfield  {author} {\bibinfo {author} {\bibfnamefont {X.}~\bibnamefont
  {Mi}}, \bibinfo {author} {\bibfnamefont {S.}~\bibnamefont {Kohler}}, \ and\
  \bibinfo {author} {\bibfnamefont {J.~R.}\ \bibnamefont {Petta}},\ }\bibfield
  {title} {\enquote {\bibinfo {title} {Electrically protected valley-orbit
  qubits in silicon},}\ }\href@noop {} {\bibfield  {journal} {\bibinfo
  {journal} {arXiv:1805.04545}\ } (\bibinfo {year} {2018})}\BibitemShut
  {NoStop}%
\bibitem [{\citenamefont {Rozhkov}\ \emph {et~al.}(2017)\citenamefont
  {Rozhkov}, \citenamefont {Rakhmanov}, \citenamefont {Sboychakov},
  \citenamefont {Kugel},\ and\ \citenamefont {Nori}}]{Rozhkov17}%
  \BibitemOpen
  \bibfield  {author} {\bibinfo {author} {\bibfnamefont {A.~V.}\ \bibnamefont
  {Rozhkov}}, \bibinfo {author} {\bibfnamefont {A.~L.}\ \bibnamefont
  {Rakhmanov}}, \bibinfo {author} {\bibfnamefont {A.~O.}\ \bibnamefont
  {Sboychakov}}, \bibinfo {author} {\bibfnamefont {K.~I.}\ \bibnamefont
  {Kugel}}, \ and\ \bibinfo {author} {\bibfnamefont {F.}~\bibnamefont {Nori}},\
  }\bibfield  {title} {\enquote {\bibinfo {title} {Spin-valley half-metal as a
  prospective material for spin valleytronics},}\ }\href@noop {} {\bibfield
  {journal} {\bibinfo  {journal} {Phys. Rev. Lett.}\ }\textbf {\bibinfo
  {volume} {119}},\ \bibinfo {pages} {107601} (\bibinfo {year}
  {2017})}\BibitemShut {NoStop}%
\bibitem [{\citenamefont {Qi}\ \emph {et~al.}(2017)\citenamefont {Qi},
  \citenamefont {Wu}, \citenamefont {Ward}, \citenamefont {Prance},
  \citenamefont {Kim}, \citenamefont {Gamble}, \citenamefont {Mohr},
  \citenamefont {Shi}, \citenamefont {Savage}, \citenamefont {Lagally},
  \citenamefont {Eriksson}, \citenamefont {Friesen}, \citenamefont
  {Coppersmith},\ and\ \citenamefont {Vavilov}}]{Qi17}%
  \BibitemOpen
  \bibfield  {author} {\bibinfo {author} {\bibfnamefont {Z.}~\bibnamefont
  {Qi}}, \bibinfo {author} {\bibfnamefont {X.}~\bibnamefont {Wu}}, \bibinfo
  {author} {\bibfnamefont {D.~R.}\ \bibnamefont {Ward}}, \bibinfo {author}
  {\bibfnamefont {J.~R.}\ \bibnamefont {Prance}}, \bibinfo {author}
  {\bibfnamefont {D.}~\bibnamefont {Kim}}, \bibinfo {author} {\bibfnamefont
  {J.~K.}\ \bibnamefont {Gamble}}, \bibinfo {author} {\bibfnamefont {R.~T.}\
  \bibnamefont {Mohr}}, \bibinfo {author} {\bibfnamefont {Z.}~\bibnamefont
  {Shi}}, \bibinfo {author} {\bibfnamefont {D.~E.}\ \bibnamefont {Savage}},
  \bibinfo {author} {\bibfnamefont {M.~G.}\ \bibnamefont {Lagally}}, \bibinfo
  {author} {\bibfnamefont {M.~A.}\ \bibnamefont {Eriksson}}, \bibinfo {author}
  {\bibfnamefont {M.}~\bibnamefont {Friesen}}, \bibinfo {author} {\bibfnamefont
  {S.~N.}\ \bibnamefont {Coppersmith}}, \ and\ \bibinfo {author} {\bibfnamefont
  {M.~G.}\ \bibnamefont {Vavilov}},\ }\bibfield  {title} {\enquote {\bibinfo
  {title} {Effects of charge noise on a pulse-gated singlet-triplet qubit},}\
  }\href@noop {} {\bibfield  {journal} {\bibinfo  {journal} {Phys. Rev. B}\
  }\textbf {\bibinfo {volume} {96}},\ \bibinfo {pages} {115305} (\bibinfo
  {year} {2017})}\BibitemShut {NoStop}%
\bibitem [{\citenamefont {Pietik\"ainen}\ \emph {et~al.}(2018)\citenamefont
  {Pietik\"ainen}, \citenamefont {Danilin}, \citenamefont {Kumar},
  \citenamefont {Tuorila},\ and\ \citenamefont {Paraoanu}}]{Pietikainen17b}%
  \BibitemOpen
  \bibfield  {author} {\bibinfo {author} {\bibfnamefont {I.}~\bibnamefont
  {Pietik\"ainen}}, \bibinfo {author} {\bibfnamefont {S.}~\bibnamefont
  {Danilin}}, \bibinfo {author} {\bibfnamefont {K.~S.}\ \bibnamefont {Kumar}},
  \bibinfo {author} {\bibfnamefont {J.}~\bibnamefont {Tuorila}}, \ and\
  \bibinfo {author} {\bibfnamefont {G.~S.}\ \bibnamefont {Paraoanu}},\
  }\bibfield  {title} {\enquote {\bibinfo {title} {Multilevel effects in a
  driven generalized {Rabi} model},}\ }\href@noop {} {\bibfield  {journal}
  {\bibinfo  {journal} {J. Low. Temp. Phys.}\ }\textbf {\bibinfo {volume}
  {191}},\ \bibinfo {pages} {354} (\bibinfo {year} {2018})}\BibitemShut
  {NoStop}%
\bibitem [{\citenamefont {Ashhab}\ \emph {et~al.}(2007)\citenamefont {Ashhab},
  \citenamefont {Johansson}, \citenamefont {Zagoskin},\ and\ \citenamefont
  {Nori}}]{Ashhab07}%
  \BibitemOpen
  \bibfield  {author} {\bibinfo {author} {\bibfnamefont {S.}~\bibnamefont
  {Ashhab}}, \bibinfo {author} {\bibfnamefont {J.~R.}\ \bibnamefont
  {Johansson}}, \bibinfo {author} {\bibfnamefont {A.~M.}\ \bibnamefont
  {Zagoskin}}, \ and\ \bibinfo {author} {\bibfnamefont {F.}~\bibnamefont
  {Nori}},\ }\bibfield  {title} {\enquote {\bibinfo {title} {Two-level systems
  driven by large-amplitude fields},}\ }\href@noop {} {\bibfield  {journal}
  {\bibinfo  {journal} {Phys. Rev. A}\ }\textbf {\bibinfo {volume} {75}},\
  \bibinfo {pages} {063414} (\bibinfo {year} {2007})}\BibitemShut {NoStop}%
\bibitem [{\citenamefont {Shevchenko}\ \emph {et~al.}(2012)\citenamefont
  {Shevchenko}, \citenamefont {Ashhab},\ and\ \citenamefont
  {Nori}}]{Shevchenko12}%
  \BibitemOpen
  \bibfield  {author} {\bibinfo {author} {\bibfnamefont {S.~N.}\ \bibnamefont
  {Shevchenko}}, \bibinfo {author} {\bibfnamefont {S.}~\bibnamefont {Ashhab}},
  \ and\ \bibinfo {author} {\bibfnamefont {F.}~\bibnamefont {Nori}},\
  }\bibfield  {title} {\enquote {\bibinfo {title} {Inverse
  {L}andau-{Z}ener-{S}t{\"u}ckelberg problem for qubit-resonator systems},}\
  }\href@noop {} {\bibfield  {journal} {\bibinfo  {journal} {Phys. Rev. B}\
  }\textbf {\bibinfo {volume} {85}},\ \bibinfo {pages} {094502} (\bibinfo
  {year} {2012})}\BibitemShut {NoStop}%
\bibitem [{\citenamefont {Denisenko}\ \emph {et~al.}(2010)\citenamefont
  {Denisenko}, \citenamefont {Satanin}, \citenamefont {Ashhab},\ and\
  \citenamefont {Nori}}]{Denisenko10}%
  \BibitemOpen
  \bibfield  {author} {\bibinfo {author} {\bibfnamefont {M.~V.}\ \bibnamefont
  {Denisenko}}, \bibinfo {author} {\bibfnamefont {A.~M.}\ \bibnamefont
  {Satanin}}, \bibinfo {author} {\bibfnamefont {S.}~\bibnamefont {Ashhab}}, \
  and\ \bibinfo {author} {\bibfnamefont {F.}~\bibnamefont {Nori}},\ }\bibfield
  {title} {\enquote {\bibinfo {title} {Dynamics of interacting qubits in a
  strong alternating electromagnetic field},}\ }\href@noop {} {\bibfield
  {journal} {\bibinfo  {journal} {Phys. Solid State}\ }\textbf {\bibinfo
  {volume} {52}},\ \bibinfo {pages} {2281--2286} (\bibinfo {year}
  {2010})}\BibitemShut {NoStop}%
\bibitem [{\citenamefont {Temchenko}\ \emph {et~al.}(2011)\citenamefont
  {Temchenko}, \citenamefont {Shevchenko},\ and\ \citenamefont
  {Omelyanchouk}}]{Temchenko11}%
  \BibitemOpen
  \bibfield  {author} {\bibinfo {author} {\bibfnamefont {E.~A.}\ \bibnamefont
  {Temchenko}}, \bibinfo {author} {\bibfnamefont {S.~N.}\ \bibnamefont
  {Shevchenko}}, \ and\ \bibinfo {author} {\bibfnamefont {A.~N.}\ \bibnamefont
  {Omelyanchouk}},\ }\bibfield  {title} {\enquote {\bibinfo {title}
  {Dissipative dynamics of a two-qubit system: Four-level lasing},}\
  }\href@noop {} {\bibfield  {journal} {\bibinfo  {journal} {Phys. Rev. B}\
  }\textbf {\bibinfo {volume} {83}},\ \bibinfo {pages} {144507} (\bibinfo
  {year} {2011})}\BibitemShut {NoStop}%
\bibitem [{\citenamefont {Gramajo}\ \emph {et~al.}(2017)\citenamefont
  {Gramajo}, \citenamefont {Dominguez},\ and\ \citenamefont
  {Sanchez}}]{Gramajo17}%
  \BibitemOpen
  \bibfield  {author} {\bibinfo {author} {\bibfnamefont {A.~L.}\ \bibnamefont
  {Gramajo}}, \bibinfo {author} {\bibfnamefont {D.}~\bibnamefont {Dominguez}},
  \ and\ \bibinfo {author} {\bibfnamefont {M.~J.}\ \bibnamefont {Sanchez}},\
  }\bibfield  {title} {\enquote {\bibinfo {title} {Entanglement generation
  through the interplay of harmonic driving and interaction in coupled
  superconducting qubits},}\ }\href@noop {} {\bibfield  {journal} {\bibinfo
  {journal} {Eur. Phys. J. B}\ }\textbf {\bibinfo {volume} {90}},\ \bibinfo
  {pages} {255} (\bibinfo {year} {2017})}\BibitemShut {NoStop}%
\bibitem [{\citenamefont {Shevchenko}\ \emph {et~al.}(2014)\citenamefont
  {Shevchenko}, \citenamefont {Oelsner}, \citenamefont {Greenberg},
  \citenamefont {Macha}, \citenamefont {Karpov}, \citenamefont {Grajcar},
  \citenamefont {H\"ubner}, \citenamefont {Omelyanchouk},\ and\ \citenamefont
  {Il'ichev}}]{Shevchenko14}%
  \BibitemOpen
  \bibfield  {author} {\bibinfo {author} {\bibfnamefont {S.~N.}\ \bibnamefont
  {Shevchenko}}, \bibinfo {author} {\bibfnamefont {G.}~\bibnamefont {Oelsner}},
  \bibinfo {author} {\bibfnamefont {Y.~S.}\ \bibnamefont {Greenberg}}, \bibinfo
  {author} {\bibfnamefont {P.}~\bibnamefont {Macha}}, \bibinfo {author}
  {\bibfnamefont {D.~S.}\ \bibnamefont {Karpov}}, \bibinfo {author}
  {\bibfnamefont {M.}~\bibnamefont {Grajcar}}, \bibinfo {author} {\bibfnamefont
  {U.}~\bibnamefont {H\"ubner}}, \bibinfo {author} {\bibfnamefont {A.~N.}\
  \bibnamefont {Omelyanchouk}}, \ and\ \bibinfo {author} {\bibfnamefont
  {E.}~\bibnamefont {Il'ichev}},\ }\bibfield  {title} {\enquote {\bibinfo
  {title} {Amplification and attenuation of a probe signal by doubly dressed
  states},}\ }\href@noop {} {\bibfield  {journal} {\bibinfo  {journal} {Phys.
  Rev. B}\ }\textbf {\bibinfo {volume} {89}},\ \bibinfo {pages} {184504}
  (\bibinfo {year} {2014})}\BibitemShut {NoStop}%
\end{thebibliography}%

\end{document}